%% file: 0-main.tex
\documentclass[manuscript,screen,nonacm]{acmart}

\usepackage{amsmath,amsfonts}
\usepackage{algorithmic}
\usepackage{bm}
\usepackage{bigstrut}
\usepackage{etoolbox}
\usepackage{graphicx}
\usepackage{textcomp}
\usepackage{xcolor}
\usepackage{xspace}
\usepackage{enumitem}
\usepackage{multicol}
\usepackage{multirow}
\usepackage{mathtools}
\usepackage{url}
\usepackage{listings}
\usepackage[utf8]{inputenc}
\usepackage{subcaption}
\usepackage{gensymb}
\usepackage{makecell}
\usepackage{hyperref}
\usepackage{soul}
\usepackage{tikz}
\usepackage[ruled,vlined]{algorithm2e}
\usepackage{pgf-umlsd}
\usepackage{balance}
\usepackage{float}
\usepackage{tabularx}
\usepackage{rotating}
\usepackage[para]{threeparttable}

\input{0-new-commands.tex}


\begin{document}

\title{Machine Learning with Confidential Computing: A Systematization of Knowledge}

\author{Fan Mo}
\affiliation{%
  \institution{Imperial College London}
  \city{London}
  \country{UK}
}

\author{Zahra Tarkhani}
\affiliation{%
  \institution{Microsoft Research}
  \city{Cambridge}
  \country{UK}
}
\author{Hamed Haddadi}
\affiliation{%
  \institution{Imperial College London}
  \city{London}
  \country{UK}
}

\input{0-abstract}
\maketitle

\input{1-introduction}
\input{2-ml}
\input{3-cc}
\input{4-cc-for-ml}
\input{5-cc-for-ml-integrity}

\input{6-tee-limitation}

\input{7-beyond}
\input{8-conclusion}

\bibliographystyle{IEEEtran}
\bibliography{reference}
\end{document}

%% file: 0-new-commands.tex
\newcommand{\mysub}[1]{\vspace{2pt} \noindent \textbf{#1}}

\newcommand{\etal}{\emph{et~al.}}
\newcommand{\ie}{\emph{i.e.,}}
\newcommand{\eg}{\emph{e.g.,}}

\newcommand*\emptycirc[1][0.8ex]{\tikz\draw[thick] (0,0) circle (#1);} 
\newcommand*\halfcirc[1][0.8ex]{%
  \begin{tikzpicture}
  \draw[fill] (0,0)-- (90:#1) arc (90:270:#1) -- cycle ;
  \draw[thick] (0,0) circle (#1);
  \end{tikzpicture}}
\newcommand*\fullcirc[1][0.8ex]{\tikz\fill (0,0) circle (#1);} 

%% file: 0-abstract.tex
\begin{abstract}

Privacy and security challenges in Machine Learning (ML) have become increasingly severe, along with ML's pervasive development and the recent demonstration of large attack surfaces. 
As a mature system-oriented approach, Confidential Computing has been utilized in both academia and industry to mitigate privacy and security issues in various ML scenarios. 
In this paper, the conjunction between ML and Confidential Computing is investigated. We systematize the prior work on Confidential Computing-assisted ML techniques that provide \textit{i})~\emph{confidentiality guarantees} and \textit{ii})~\emph{integrity assurances}, and discuss their advanced features and drawbacks. Key challenges are further identified, and we provide dedicated analyses of the \emph{limitations} in existing \emph{Trusted Execution Environment} (TEE) systems for ML use cases. 
Finally, prospective works are discussed, including grounded privacy definitions for closed-loop protection, partitioned executions of efficient ML, dedicated TEE-assisted designs for ML, TEE-aware ML, and ML full pipeline guarantees. By providing these potential solutions in our systematization of knowledge, we aim to build the bridge to help achieve a much stronger TEE-enabled ML for privacy guarantees without introducing computation and system costs. 
\end{abstract}

%% file: 1-introduction.tex
\section{Introduction}

Machine learning (ML) has been established as the most promising approach to learning patterns from data, and its applications exist across data processing of online surfing~\cite{boselli2018classifying,naumov2019deep}, financial data~\cite{leo2019machine}, health care~\cite{ahmad2018interpretable,shailaja2018machine}, autonomous cars~\cite{meyer2019deep,ravindran2020multi}, and almost every data-driven field around us. 
The wide applications have also led ML models, specifically Deep Neural Networks (DNNs), to be run on a diverse set of devices -- from low-end IoT/mobile devices to high-performance cloud/data centers to provide both ML model training and inference services. 
However, the wide applications also open a large attack surface on ML's security and privacy due to a large number of mistrusting entities and complex software/hardware infrastructure involved. 
Recent research has massively explored the way to attack ML. Several examples are \eg~model stealing~\cite{orekondy2019knockoff,tramer2016stealing,shen2022model,rakin2022deepsteal}, model inversion~\cite{fredrikson2015model,he2019model}, model/data poisoning~\cite{fang2020local,chen2017targeted,shejwalkar2022back}, data reconstruction~\cite{zhu2019deep,yin2021see}, membership/attribute inference~\cite{shokri2017membership,melis2019exploiting,carlini2022membership}, etc. 
Vulnerabilities shown by these attacks have increasingly perturbed ML developers and users due to the negative consequences they can cause: the destruction of their ML training process, the leakage of private information, or the infringement of intellectual property.

Specifically, ML faces security and privacy issues mainly due to, firstly, the \emph{complexity of pipelines} that involve many system/software stacks to provide modern features like parallelization and acceleration.
A full ML pipeline covers the collecting of raw data, the complete training and inference phases, the later use of the trained ML model for prediction, and the potential re-train and re-use of the ML model.
Because the involved participants \ie~data owners, the host of ML computation, the model owners, and the result receivers are most likely to be different entities, the pipeline can be segmented across mutually mistrusting individuals and, consequently, is left with a broad attack surface.
Second, the \emph{weak robustness and low interpretability} of ML algorithms aggravate the security and privacy problems.
The modern ML model (\ie~DNN) is a large-scale nonlinear system that has been progressively adjusted using stochastic gradient descent (SGD) during training. Until now, there is still a lack of an accurate representation/description of the dynamics of this training progress, and it has been increasingly hard in terms of model explainability along with the evolving of complicated training techniques and the size of large models (\eg~large language models GPT-4 or Llama).
Consequently, ML models often have relatively weak robustness, which has been shown in adversarial examples~\cite{goodfellow2014explaining} or poisoning attacks~\cite{chen2017targeted}), \ie~a small malicious change to the training process or input data could cause large negative impacts that are difficult, or even impossible, to detect.

Among many candidates to achieve trustworthy ML, a system-oriented solution that provides a trusted environment within the system, called Confidential Computing, has emerged as one promising approach because of its high performance.
Confidential Computing\footnote{There are other cryptography-based techniques such as Multi-party Computation and (Leveled or Full) Homomorphic Encryption used for guaranteeing ML confidentiality and integrity; they are categorized as Privacy-preserving Computation to distinguish with TEE-based Confidential Computing~\cite{graepel2012ml, big2019handbook, choquette2021capc, ccc}.}, defined in the Linux Foundation project Confidential Computing Consortium's whitepapers~\cite{cctechnical, ccapplication}, is the protection of data in use, in addition to data protection in storage and transmission, by utilizing hardware-based Trusted Execution Environments (TEEs).
TEE is one of the emerging techniques in enabling isolated and verifiable code execution inside protected memory (sometimes called enclaves or secure world depending on realization and manufacturers), separated from the host privileged system stacks such as operating system or hypervisor. 
Both hardware and software changes are typically required to achieve proper isolation on modern processors (\ie~CPUs) at an adequate trustworthiness level.
Compared to the conventional external secure processor, like the Trusted Platform Module that uses separated hardware from the rest of the board/SoC, TEEs still remain and run on existing processors. This delivers the flexibility in switching contents between normal execution mode and trusted execution mode and further provides much more computational resources in the system.
Because of their flexibility and reasonable performance in supplying trustworthiness, TEEs have become nearly an essential component of modern processors, and various types of TEEs are implemented by \eg~ARM, Intel, AMD, and NVIDIA to realize Confidential Computing.

The rapid growth of Confidential Computing pushes a line of research work that leverages this technique for trustworthy ML computations~\cite{ohrimenko2016oblivious, tramer2018slalom, hunt2018chiron, lee2019occlumency,mo2019efficient, mo2021ppfl, mo2022privacy}, and a great number of commercial products are available nowadays, \eg~Confidential Computing service from Microsoft Azure~\cite{ccazure}, Amazon Web Services~\cite{ccnitroenclaves}, Google Cloud~\cite{ccgooglecloud}, etc. 
Although many attempts have been made to combine Confidential Computing and ML and several limitations (\eg~still limited TEE resources for ML use) are summarized~\cite{duy2021confidential}, there is still no systematization of Confidential Computing-assisted ML, challenges, and especially potential solutions in a long term vision.
There are still unclarities of what the critical scientific and engineering obstacles are and how to overcome them to achieve trustworthy ML.
Furthermore, both Confidential Computing and ML have complicated mechanisms respectively; a lack of such systematized analysis makes a sophisticated combination of these two impossible, which even further harms both the integrity of Confidential Computing and the performance of ML.

To alleviate the security and privacy problems in ML, in this paper, we present a systematic analysis of Confidential Computing-assisted solutions.

\mysub{Organization.}
The next sections of the paper are organized as: 
Section~\ref{sec:ml} focuses on ML, and it describes the pipeline and paradigms of ML and existing attack surfaces; 
Section~\ref{sec:cc} focuses on Confidential Computing, and it mainly outlines the threat model and key components of Confidential Computing;
Then, Section~\ref{sec:cc_for_ml} summarizes challenges and existing solutions when using Confidential Computing to provide trustworthy ML, specifically in terms of confidentiality;
Section~\ref{sec:cc_for_ml_integrity} further presents the challenges and solutions of guaranteeing ML \emph{integrity} with Confidential Computing;
Section~\ref{sec:limit_tee} outlines the existing TEE limits in Confidential Computing that hinder its usage in ML;
Section~\ref{sec:beyond} concludes the current research stage and further work.

%% file: 2-ml.tex
\section{Machine Learning}
\label{sec:ml}

Machine learning (ML) automates pattern learning from large datasets in the sense that analysts do not need to manually explore data's latent features and their correlations~\cite{jordan2015machine, lecun2015deep}. In this section, we look into ML pipelines, common paradigms, and attack surfaces that exist among them.

\subsection{Pipeline}
\begin{figure}
    \centering
    \includegraphics[width=0.95\columnwidth]{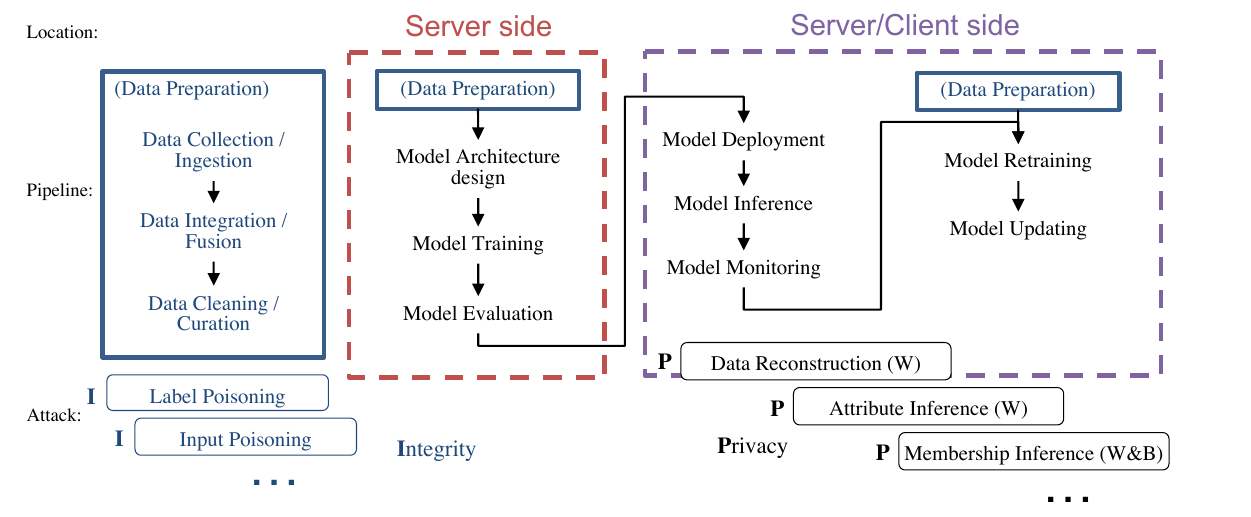}
    \caption{The schematic diagram of the machine learning pipeline and resided attack surfaces across the pipeline. `I' refers to Integrity-related attacks. `P' refers to Privacy-related attacks. `W' refers to that one attack is generally in white-box form, while `B' refers to black-box form.}
    \label{fig:ml_pipeline}
\end{figure}

The ML pipeline codifies and automates the workflow to produce and apply ML algorithms or models. It consists of multiple sequential steps forming the lifecycle of using ML. We present the general ML pipeline in Figure~\ref{fig:ml_pipeline}. 
The form can change slightly depending on different ML use cases but the core parts are interchangeable overall as identified by works from Microsoft~\cite{amershi2019software}, IBM~\cite{ibmallifecycle}, and ETSI~\cite{etsigrsai}. Key stages of the ML pipeline are given below.

\mysub{Data preparation.}
Data are the basis for ML evidently because ML requires tons of inputs to attempt ``trial and error'' and learn the patterns. 
In supervised learning~\cite{lecun2015deep}, all data are labeled in some ways, which conventionally need extensive efforts of human annotation. 
In semi-/un-supervised learning~\cite{zhu2009introduction, berthelot2019mixmatch, le2013building}, parts/all of the data are unlabeled due to the difficulty of labeling. In some cases, unlabelled parts of data can be ``self-learned'' after the model already gains a certain level of prediction ability on these data~\cite{van2020survey}. Also, no label is needed when the model aims at learning encoding features and then decoding features back to original data (\eg~Autoencoder~\cite{tschannen2018recent, vahdat2020nvae}).
In reinforcement learning~\cite{sutton2018reinforcement}, data are in the form of sequences of action, observations, and rewards, which are sequentially produced/updated during the learning process of a policy function.
Regardless of the form, ML is always \emph{data-hungry}.
Thus, many efforts have to be made to obtain a fair amount of data, to cover possible features that the data have in their distribution space.
One common technique to increase the amount of data is data augmentation which adds slightly modified copies or synthetic copies of existing data~\cite{shorten2019survey, xie2020unsupervised}.
In addition, another critical issue is that the collected dataset can be very noisy or contain bad records/features from which the ML model should not learn. Then, it is almost always necessary to apply filter-like algorithms (\eg~data clean or data selection schemes) to refine the data or select training data by measuring the importance of data samples~\cite{evans2018learning, katharopoulos2018not, mindermann2022prioritized}. After that, collected data are typically partitioned as the training set and the test set (sometimes additionally, the validation set) for training and evaluating the ML model.

\mysub{Model training.}
Most model training aims to acquire a function $f_{\theta}(x)$ for $\theta \in \Theta$, capable of mapping an input $x$ to a predicted decision $\hat{y} = f_{\theta}(x)$ near to the true decision $y$, where $\theta$ and $\Theta$ denotes model parameters and its corresponding parameter space.
The form of $f_{\theta}(\cdot)$ other than ${\theta}$ and implicitly included in function $f()$ is called model hyperparameters; specifically regarding DNNs, hyperparameters are determined by chosen DNN architectures \eg~the type of layers, the number of neurons, etc. Note that the underlying hyperparameters also affect the training performance (sometimes a lot), which can be tuned manually or automatically using neural architecture search~\cite{elsken2019neural}.
Nevertheless, model training, conventionally, regards the searching for the best $\theta$ set that is achieved by minimizing the loss $\ell(y, f_{\theta}(x))$ using SGD methods~\cite{lecun1998gradient, bottou2010large} on training sets, \ie~stepped descent by $\theta \xleftarrow{} \theta - \eta \nabla_\theta f_{\theta}(x)$, where $\eta$ is the step size multiplier known as the learning rate.
To update $\theta$ of the complete DNN model, the loss is backward passed through the DNN starting from the last layer to the first, which is called \emph{backward propagation}.
After reaching a certain level of performance on the validation set (or the other subset apart from the training set in cross-validation~\cite{kohavi1995study, arlot2010survey}), the model training stops. Finally, $f_{\theta}(\cdot)$'s final model utility/accuracy is reported on the test set.

\mysub{Model deployment/Inference.}
With the above trained DNN model, one can feed unseen input data $x_{\text{new}}$ into the DNN model to get the prediction $\hat{y} = f_{\theta}(x_{\text{new}})$. 
This is called the inference stage. 
It is worth noting that the inference could happen at both the \emph{model provider} side or the \emph{data owner} side depending on the trust and collaboration form between the providers and the users. 
For instance, in typical cases of ML inference, the model needs to be deployed i)~on a server-side device to support centralized service \eg~ML inference as a service, or ii)~on the users' devices for distributed local inference.

\mysub{Retraining/Updates.}
To adapt to data with unseen features over time, ML models require retraining/updating their parameters.
Such updates can be achieved by fine-tuning the model with newly collected data in some forms. Especially, the fine-tuning could focus on the last layers of the model, which is one way to achieve transfer learning~\cite{pan2009survey, marcelino2018transfer}, as the first layers already generalize well on one type of data (\eg~images in general)~\cite{pan2009survey, yosinski2014transferable, zeiler2014visualizing} in order to improve the training efficiency. 
Two examples of such a technique are i)~back-propagation on the last layers only~\cite{marcelino2018transfer} and ii)~low-shot learning with weights imprinting on the last layer~\cite{qi2018low}. 
The reason for updating the last layers preferentially is related to one of DNN's learning behaviors; its first layers tend to learn more general features that all input data should have, while the last layers learn more specific features that do not cover some new incoming data~\cite{pan2009survey, yosinski2014transferable, zeiler2014visualizing}.
However, whether we can take such a ``shortcut'' or not depends on how new the data's features are with respect to the pre-trained DNN model.
Similarly to model inference, this stage can happen on not only the server side but also the client side depending on the cases. Retraining on the server usually requires newly collected data that may be from clients.
In the other case, after having an updated model on the client side, it can further share it with other users or the central server for knowledge sharing, sometimes called distributed learning or federated learning~\cite{mcmahan2017communication, li2020federated, kairouz2019advances}.

\subsection{Paradigm}
\label{sec:ml_paradigm}
The stages of the ML pipeline are typically distributed among different participating entities.
Such distribution of ML process is partially due to the increasing privacy/security concern raised by sharing data, as well as network/computation-resource constraints, and therefore, aims at a better trade-off among model utility, privacy preservation, and system performance and communication cost. 
Based on the location of the main ML computation (\eg~training or inference), we categorize them based on the well-accepted concept: \emph{centralized machine learning} and \emph{distributed/federated machine learning}.

\mysub{Centralized machine learning.}
In centralized ML, \emph{all} of the training or inference computations are conducted at a central place such as a server~\cite{abadi2016tensorflow, yao2017complexity}.
For both training and inference, the data have to be collected and stored at/near the center, and then the ML algorithm performs training or inference functions on the data. In such a case, potential retraining will also happen on the server with newly collected data. 
Centralized ML usually achieves the best training performance compared to all forms of federated ML, because it has all data on hard with the probability of attempting various training strategies that are impossible when data are distributed.
Nevertheless, centralized ML raises privacy concerns about others' data, considering that all these data are required to be gathered together. Therefore, centralized training/inference is usually chosen for the cases where collected data are public (or non-private), or the data contributors are collaborators or have \textit{some} form of trust/contracts with the central server.

\mysub{Distributed machine learning.}
In the other cases, when data owners do not have the willingness to share data, federated ML is one potential approach, where the \emph{ML model}, instead of the private data, is shared among participating entities~\cite{li2014scaling, mcmahan2017communication}. 
Such a ML paradigm enables the client to perform computations on the model and collaborate with other clients with some specifically designed protocols~\cite{so2020scalable, verbraeken2020survey}.
Specifically, in terms of \emph{ML inference} in a distributed way, it is quite straightforward that the model owner \emph{distributes} its ML model to end-users, so that users can perform inference locally on their devices.
Potentially, they can further conduct fine-tuning on the model for personalization but still for their own usage.
Regarding the \emph{ML training}, the terminology `federated learning (FL)'~\cite{mcmahan2017communication, li2020federated, kairouz2019advances} has been intensively used referring to the paradigm that users send the locally updated model parameters to the server without the need of revealing their data.
User's updated models are gathered and then aggregated as one global model on the server, using chosen aggregation techniques \eg~FedProx~\cite{fedprox}, FedAdam~\cite{reddi2020adaptive}, FedMA~\cite{wang2020federated}, etc.
Unfortunately, despite that such FL designs never share local data, recent works demonstrate the probability of
retrieving some form of sensitive information (\eg~attributes of the local data) from the model parameters themselves~\cite{geiping2020inverting, melis2019exploiting, zhu2019deep}. However, recent research reveals that even in federated learning, the shared model parameters still make the system vulnerable due to information leakage and potentially malicious changes in model parameters. The taxonomy of ML paradigms is further clarified when elaborately describing the underlying protection scenarios in Section~\ref{sec:cc_ml_overview} and Figure~\ref{fig:protection_classes}.

\subsection{Attack Surface}
\label{subsec:attack_surface}
The attack surface can cover the complete pipeline caused by the potential multiple vulnerabilities mentioned above. 
Here, we categorize existing attack vectors based on the core underpinning of information security into i)~\emph{confidentiality}, \ie~privacy of data and intellectual property of models, ii)~\emph{integrity} of the ML process. 
Note that while the full CIA triad includes Confidentiality, Integrity, and Availability, the attacks that specifically aim at ML system's Availability, \eg~Denial-of-Service (DoS) attacks, have not drawn much attention from both industry and academia currently, probably due to the high similarity of availability problems between conventional systems and ML systems. Further note that despite their differences, we use \emph{confidentiality} and \emph{privacy} interchangeably sometimes considering that the confidentiality of the data or model usually represents the privacy of their owners.
Thus, the attacks on every vulnerability in the ML pipeline are viewed in light of confidentiality and/or integrity.

\mysub{Confidentiality-related attacks.}
An adversary can be honest-but-curious, \ie~honestly performing ML training/inference without changing the computation results, but actively explore (unauthorized) sensitive information of data and the model.
Such attack vectors exist across stages of the ML pipeline including model training, model deployment/inference, or model updates because the data and models are typically owned by multiple mistrusting entities.
In terms of centralized ML, the computation host can directly access incoming data~\cite{ribeiro2015mlaas, hesamifard2018privacy} without any protection schemes, thus violating the privacy of data.
In federated ML, the host who orchestrates all clients' local training can access their updated models and further infer private information about their local data using these models~\cite{geiping2020inverting, melis2019exploiting, nasr2019comprehensive}, although it cannot directly access clients' private data. 
Common attacks on private data can be categorized into i)~the Data Reconstruction Attack (\textbf{DRA})~\cite{zhu2019deep, geiping2020inverting, hitaj2017deep}, aiming at reconstructing original input data based on the observed model or its gradients, ii)~ the Attribute Inference Attack (\textbf{AIA})~\cite{melis2019exploiting, jia2018attriguard}, aiming at inferring the value of users' private properties in the training data, and iii)~the Membership Inference Attack (\textbf{MIA})~\cite{nasr2019comprehensive, shokri2017membership}, aiming at learning whether specific data instances are present in the training dataset.
These three categories thoroughly cover the spectrum of privacy-related attacks. DRAs lead to complete privacy leakage about target data, while MIAs are the mildest with leakage of data's membership information only. AIAs broadly cover middle levels of privacy leakages.

On the other side, the confidentiality of models is also targeted by attacks.
One model owner usually considers a model as its intellectual property as the model has been trained with great effect (\eg~computation power, data collection, and data cleaning work), such as the pre-trained model in machine learning as a service (MLaaS) hosted by Amazon Web Services (AWS). 
One well-explored type of attack, called the Model Stealing Attack (\textbf{MSA}), aims to counterfeit the functionality of a model by exploiting black-box access in ML inference~\cite{orekondy2019knockoff, tramer2016stealing}. One example is to query a large number of prediction results and then use these input-output pairs to solve ``equations" or ``retrain'' a model somehow~\cite{yu2020cloudleak, tramer2016stealing}. This problem becomes severe when the model size is smaller and sometimes related to highly valuable training data that are hard to obtain.

\mysub{Integrity-related attacks.}
In addition, attackers can actively exploit training/inference results, thus, violating the integrity of the ML process.
The fairly most well-known adversarial attack, called Adversarial Examples (\textbf{AE})~\cite{goodfellow2014explaining, moosavi2017universal}, adds calibrated noises to one image/audio, consequently leading to wrong prediction results during inference. The input with noisy perturbation is invisible to human beings making such attacks really hard to detect.
In a more influential case, attackers take full control of wearable BCI (Brain computing interface) devices (\eg~wheelchair, robotic arms, or drone) by compromising input signals to on-device ML models~\cite{tarkhani2022enhancing}, which sometimes can be life-threatening.
Besides, previous work has also shown the possibility of model accuracy degradation by only compromising thread scheduling on multi-threaded ML pipeline~\cite{sanchez2020game}. We note that although these attacks manipulate input data, they break the integrity of ML inference.

Regarding ML training, specifically in federated learning, Model Poisoning Attacks (\textbf{MPA}) can lead to Byzantine fault~\cite{fang2020local}. Such faults easily occur because the server does not control clients' local training and cannot verify whether their behaviors have followed the promised training processes.
For instance, if one error in values of locally updated parameters has been aggregated into the global model, the complete global model can become unusable or even fail to converge.
In a more sophisticated way, one can manipulate the training set \eg~by adding data with calibrated noisy labels~\cite{tolpegin2020data} to drive the training process into undesired ways, namely Backdoor Attacks or Data Poisoning Attacks (\textbf{DPA}) in some works.
This type of attack fools the classifier into having wrong boundaries for some specifically chosen data points. The attackers/other users will be able to trigger such backdoors in later use after model deployment~\cite{liu2020reflection, bagdasaryan2020backdoor, jia2022badencoder}.

\mysub{White-box or black-box.}
Here we further describe a common way of categorizing ML attacks: white-box and black-box. Naturally, adversaries have different levels of prior information before performing attacks. Here, ML attack surfaces can be categorized based on whether one attack requires access to the internal architecture of a ML model~\cite{papernot2018sok, sablayrolles2019white}. 
All above-mentioned attacks can be grouped into white-box or black-box based on the level of required knowledge of the target ML model.
With regards to black-box attacks, model stealing usually starts from the outside without any prior information and aims to learn the model itself. Also, membership inference attacks on data privacy are typically black-box because it is already much more efficient than white-box attacks as shown in previous research~\cite{sablayrolles2019white, yeom2018privacy}. White-box access does not significantly increase the attack `advantage' in disclosing membership privacy.
On the other side, the white-box attack includes almost all data reconstruction attacks, some adversarial example attacks, and attribute inference attacks. These attacks usually require detailed model parameters/gradients in order to be performed or have reasonable performance.
We refer to Papernot~\etal's survey~\cite{papernot2018sok} and NIST's publication~\cite{vassilev2024adversarial} for more details about the taxonomy and terminology of ML's security and privacy vulnerabilities.

%% file: 3-cc.tex
\section{Confidential Computing and Trusted Execution Environments}
\label{sec:cc}

Confidential Computing~\cite{ccc} is a quickly emerging technology that leverages Trusted Execution Environments (TEE) to provide a level of assurance of privacy and integrity when executing codes on data.
Most Cloud service vendors, \eg~Google Cloud and Microsoft Azure, have now started providing Confidential Computing services, which usually use the TEE solution supplied by processor manufacturers such as ARM, Intel, AMD, and NVIDIA~\cite{russinovich2021toward}. 
In this section, we first describe threat models considered in Confidential Computing and the key components of it. Then we discuss the software stacks useful for ML application developers.

\subsection{Threat Models}
By considering a stronger adversary who has full access to the host privileged system stacks~\cite{cctechnical}, Confidential Computing significantly reduces the trusted computing base (TCB). By default, TEEs assume that the attacker controls the host OS/hypervisor as well as the host applications executing in userspace. She could, for example, access and corrupt the host OS kernel and process resources (\eg~memory, threads, network, files, and locks).
Hence, the adversary can be the service provider, the device owner itself, malicious third-party software installed on the devices, or a malicious or compromised host OS/hypervisor. 
Under such an environment, they can perform any possible white-/black-box or confidentiality-/integrity-related attacks.

Confidential Computing enables trust between mutually mistrusting components by utilizing (preferable hardware-based, attested) TEEs~\cite{whitepapertee}. Chip manufacturers usually have different realizations for their own chip structure. As an example, Arm designs its TEEs as an isolated environment running in parallel with the host OS (also known as the rich OS), allowing to securely store data and execute arbitrary code on an untrusted device, with reasonably good performance, through protected and measured memory compartments.
Thus, one can demand a remote TEE to perform computation on her (sensitive) code and/or data without revealing them to the TEE's host.
Few TEEs, \eg~AWS Nitro Cards~\cite{ccnitroenclaves}, could also extend the level of Hardware-based isolation to peripherals and persistence storage~\cite{guan2017trustshadow,bahmani2021cure} via a range of trusted PCIe interfaces or TEE-IO/device interfaces like TDISP (TEE Device Interface Security Protocol), IDE (Integrity and Data Encryption), SPDM (Secure Protocol and Data Model), which consequently By extending the trust boundary to cover more system components~\cite{kaplan2023hardware}.

The recent commoditization of TEEs both in high-end and low-end mobile devices makes it an ideal candidate to achieve confidentiality and integrity in ML. 
However, if the code inside an enclave has vulnerabilities, TEEs do not guarantee the protection of the integrity and confidentiality of that computation.
Consequently, the attacker may be able to exploit weaknesses in TEEs system~\cite{10.5555/3241189.3241231} and their interfaces/APIs in such conditions~\cite{khandaker2020coin,10.1145/3319535.3363206}.

However, TEEs are vulnerable to various side- and covert-channel attacks~\cite{Chen2021voltpillager, li2022systematic}, physical attacks~\cite{Lipp2021Platypus}, confused deputy attacks~\cite{machiry2017boomerang}, Denial-of-Service (DoS) attacks, etc~\cite{cerdeira2020sok, xia2022secret}. These attack vectors are hard to defend against in the standard TEE threat models~\cite{singh2021enclaves} and are considered out-of-scope attacks, although we briefly discuss these attacks and corresponding defenses in the discussion section.

\subsection{Key Components of Confidential Computing}

\begin{figure}
    \centering
    \includegraphics[width=0.95\columnwidth]{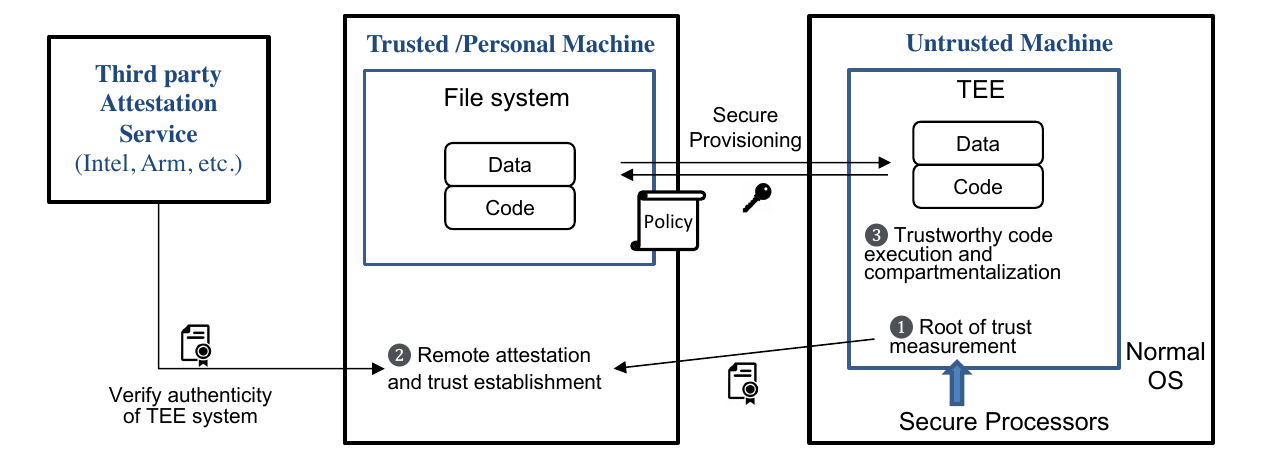}
    \caption{The schematic diagram of Confidential Computing by utilizing the Trusted Execution Environment (TEE).}
    \label{fig:cc_pipeline}
\end{figure}

\input{tables/tee-framework-sum.tex}
We give an overview of the Confidential Computing structure in Figure~\ref{fig:cc_pipeline}, then present key components of it below.

\mysub{Root of trust measurement.}
Confidential Computing provides hardware-assisted functionalities to establish trust between TEEs and the external computing environment. First, by enabling Root of Trust (RoT) measurement it ensures the integrity of the TEE system~\cite{whitepapertee, costan2016intel}. 
When the processor measures the chain of all critical system software (including boot loader, firmware, OS, hypervisor, and TEE system) before launching in-enclave code, any integrity violation of the initial state could be detected.
A compromised code in the chain cannot escape from being measured and therefore from being detected.
The measurement needs hardware unique keys which will be attested later to ensure the integrity of the TEE system~\cite{whitepapertee, costan2016intel}. 

\mysub{Remote trust establishment and attestation.}
Before deploying software or data into TEEs, one needs to remotely attest the TEE system's integrity.
Remote attestation~\cite{costan2016intel, haldar2004semantic} allows the user to determine the level of trust/integrity of a remote TEE before transmitting her sensitive data/code. 
It enables the user to authenticate the hardware, verify the trusted state of the remote TEE, and check whether the intended software is securely running inside the TEE. 
The authentication starts with signing the RoT measurement report with a private key that is securely stored on the system's hardware. The report then is sent to the attestation service, which verifies the signature using the system's public key, and checks the report against a trusted reference, \eg~a known-good configuration of the system. The attestation service then generates an attestation certificate that contains a signed statement of the system's integrity. Finally, the attestation certificate is sent to the attester.
Attestation could be conducted by a third party besides the user and host~\cite{coker2011principles} or directly between the service provider and enclave owner~\cite{scarlata2018supporting}.
For example, the attestation server could be launched by the processor manufacturer/vendor, as in Intel's SGX attestation service~\cite{johnson2016intel}, which can be different from the host service provider (\eg~a cloud provider). Nitro Enclave provided by AWS uses its own remote attestation service (\eg~AWS Key Management Service). In such a case, one AWS Nitro Enclave customer provides assurances to their downstream customer.

\mysub{Trustworthy code execution and compartmentalization.}
The core feature provided by Confidential Computing is the hardware-assisted isolation of enclaves/TEEs from the untrusted environment~\cite{whitepapertee}. In general, enclave memory regions are protected and managed through hardware and TEE's system software.
Isolation mechanisms for processor architectures can be different (\eg~for ARM, Intel, and AMD). For instance, SGX relies on Memory Encryption Engine (MEE) to protect the confidentiality, integrity, and freshness of the CPU-DRAM traffic over enclave memory ranges, while TrustZone relies on separate page tables, hardware privilege layers (\eg~EL3 and Secure EL2/EL1/EL0), and TrustZone address space controller (TZASC)\footnote{There are software/language-based TEEs and compartmentalization techniques without the needs of hardware-based guarantees (\eg~seL4), but it is not included here as they usually have lower security assurance and are not mainstream in Confidential Computing.}. 
More importantly, they run enclaves in different privilege and execution modes. For instance, SGX enclaves run in userspace mode, while AMD's SEV and Intel TDX~\cite{tdx} run them on hypervisor mode, and TrustZone secure world runs as a virtual separated core from the host core.

\mysub{Trusted IO and Device Interfaces.}
Moreover, depending on the TEE/enclave architecture and threat model, it can support additional peripheral compartmentalization and persistent storage sealing. 
In Table~\ref{tab:tee_summary} we summarize common hardware-assisted TEEs with their providers, supported processors, and features. 
Among them, TrustZone~\cite{ngabonziza2016trustzone} is developed from a very early period, and it is the most widely-used TEE on mobile/ubiquitous devices. However, it has several constraints including the inflexible protected memory, supporting only one single secure world, and restricted third-party application adaption. To overcome these issues, recently ARM proposed Confidential Compute Architecture (CCA)~\cite{li2022design, armcca} which aims to run more flexible enclaves (called realm) in parallel with TrustZone.
Therefore, modern TEEs support multi-enclaves and more flexible secure memory sizes with different TCBs.
TEE systems that include an OS-like software stack usually have a large TCB size, which provides higher usability to developers and users. For instance, when TEEs provide such an OS, they statically link applications into the in-enclave \emph{kernel}; consequently, unmodified applications could be deployed into enclaves.
In another line, CCA will be introduced in Armv9-A platforms, while billions of Armv8-A powered devices are still on the market. To tackle such a problem, IceCap, an internal project in Arm, supports a form of isolate on compatible Armv8-A devices by providing a hypervisor with a negligible, attestable TCB based on the high-assurance seL4 microkernel~\cite{mulligan2021confidential}. 
VirtCCA further implements a virtualized CCA on platforms equipped with the Secure EL2 extension available from ARMv8.4 and earlier platforms that lack Secure EL2 support with minimal overload~\cite{xu2023virtcca}.
However, as mentioned, there are still many open problems such as behavioral measurement and attestation in addition to the TEE itself.

There has also been progress on speeding up computation-heavy TEEs with new interface technologies recently. For example, SEV-TIO builds upon various industry standards such as PCIe's TDISP, IDE, SPDM, and others. It establishes a secure connection between a device and a specific SEV-SNP VM, ensuring both confidentiality and integrity of communication. Such a design allows devices to attest guest VMs, preventing interactions with malicious devices and facilitates high-performance DMA without bounce buffering, while previously, device communication relied on shared (unencrypted) memory pages, requiring data copying into specific pages before device access~\cite{sahita2023cove}.
There is also TEE/OPTEE integration of NXP's Resource Domain Controllers (RDC/XRDC/TRDC) that provide support for the isolation of peripherals and memory~\cite{opteenxp}. All these designs could achieve a more flexible/effective compute pipeline with TEEs.

\subsection{Partitioning Frameworks for Developers}

With the low-level hardware and system stacks only, it is still hard for application developers to utilize TEEs. 
Hence, depending on the use cases, various software vendors provide deployment and partitioning frameworks for running a diverse set of workloads within enclaves.
For example, more recently Amazon Nitro enclaves and Enarx support running the same binary within different types of TEEs through in-enclave WebAssembly (WASM) support, inspired by Haven~\cite{baumann2015shielding} that ports Drawbridge~\cite{porter2011rethinking} (a Windows library OS) inside an SGX enclave. 
In line with above mentioned IceCap, Veracruz is designed to further support collaborative privacy-preserving computations on multiple platforms including CCA by providing minimal filesystem, i/o, and in-enclave WebAssembly~\cite{mulligan2021confidential, brossard2023private}.
Similarly, Graphene-SGX~\cite{tsai2017graphene} ports the Linux-based Graphene library OS in an enclave. 
Scone~\cite{arnautov2016scone} tries to reduce TCB by porting \text{musl libc} and a portion of Linux Kernel Library (LKL)~\cite{sgx2019lkl}.  TrustShadow~\cite{guan2017trustshadow} also protects unmodified applications from the host OS by trapping application exceptions and system calls inside TrustZone, passing the calls to the OS, and verifying the output after parameter marshaling. Other widely used TEE-enabled deployment frameworks include confidential containers and Kubernetes services, which are especially popular and easy to use for ML workloads, as offered by cloud providers such as Azure confidential Kubernetes Service~\cite{AKS-msft} and Google Confidential GKE Nodes~\cite{GKE_2024}.

As another approach, various TEE partitioning frameworks split applications into trusted and untrusted components, \eg~Intel SGX SDK~\cite{intel2019linuxsgx}, Microsoft Open Enclave~\cite{microsoft2019openenclave}, Google’s Asylo~\cite{Google2018Asylo}, OP-TEE~\cite{optee}, and Keystone~\cite{dayeol2019keystone}.
There are also language-specific partitioning frameworks such as Civet~\cite{tsai2020civet} for porting Java classes into SGX enclaves, Trusted Language Runtime~\cite{santos2014using} for running portions of C\# applications inside TrustZone, and Glamdring~\cite{lind2017glamdring}, a compiler for partitioning applications into SGX enclaves via code annotation.
Table~\ref{tab:tee_framework} summarises primary commercial TEE frameworks.
However, not all of them are feature-rich enough to support ML use cases. Particularly, mobile vendors (\eg~TEE of Qualcomm~\cite{qualcommtee}, Trustonic~\cite{trustnictee}, or Huawei~\cite{busch2020unearthing}) also only allow for limited TEE operations to avoid security risks inside TrustZone secure world~\cite{machiry2017boomerang}. Besides, such security concerns link to the less privileged execution of enclaves as well as ease of programming. 

%% file: tables/tee-framework-sum.tex
\begin{sidewaystable*}
\vspace{440pt}
\caption{Summary of dominant hardware-assisted trusted execution environments}\label{tab:tee_summary}%
  \footnotesize
  \setlength\extrarowheight{2pt}
  \begin{threeparttable}
  \resizebox{1\textwidth}{!}{%
    \begin{tabular}{|p{9.085em}|p{3.5em}|p{8.25em}|l|p{9.585em}|p{6.665em}|l|p{6.75em}|c|p{6.335em}|}
    \hline
    \textbf{Main Stream\newline{}Hardware TEEs} & \textbf{ISA \&\newline{}Year} & \textbf{Supported \newline{}Processors} & \multicolumn{1}{p{6.415em}|}{\textbf{Number of \newline{}Isolation}} & \textbf{TCB Size (excluding \newline{}CPU/SoC package)$^\alpha$} & \textbf{Secure \newline{}Memory Size$^\delta$} & \textbf{Attestation} & \multicolumn{1}{l|}{\textbf{Application}} & \multicolumn{1}{p{4.835em}|}{\textbf{Application Unchanged}} & \multicolumn{1}{l|}{\textbf{Protections}} \bigstrut\\
    \hline
    Software Guard \newline{}Extensions (SGX) & Intel\newline{}(2013) & \multicolumn{1}{l|}{6th Intel CPU +} & multi-enclaves & small.\newline{}BIOS or firmware & up to 128MB & \multicolumn{1}{p{5.585em}|}{remote\newline{}Intel service} & \multicolumn{1}{l|}{desktop-level} & \emptycirc & confidentiality\newline{}Integrity \\
    \hline
    SGX 3rd \& 4th Gen & Intel\newline{}(2022/23) & \multicolumn{1}{l|}{Xeon Scalable CPUs} & multi-enclaves & large.\newline{}BIOS or firmware & up to 512GB \newline{} 1TB if using XTS & \multicolumn{1}{p{5.585em}|}{various remote services} & \multicolumn{1}{l|}{desktop-level} & \halfcirc & confidentiality\newline{}Integrity \\
    \hline
    \multicolumn{1}{|l|}{TrustZone (TZ)} & Arm\newline{}(2005) & ARMv6-A + / \newline{}Armv8-M + & \multicolumn{1}{p{6.415em}|}{single \newline{}secure world} & small.\newline{}firmware, TZ-kernel & typically up to \newline{}$16\sim64$MiB & \multicolumn{1}{p{5.585em}|}{fTPM-based\newline{}\cite{vtpm}} & mobile-level\newline{}desktop-level & \emptycirc & confidentiality\newline{}integrity\\
    \hline
    Secure Encrypted \newline{}Virtualization (SEV) & AMD\newline{}(2016) & AMD EPYC + & multi-VMs & large. \newline{}VM's OS, firmware & up to available \newline{}system RAM & \multicolumn{1}{p{5.585em}|}{secure \newline{}processor} & \multicolumn{1}{l|}{enterprise-level} & \fullcirc & \multicolumn{1}{l|}{confidentiality} \bigstrut\\
    \hline
    PMP-based TEE$^\beta$ & RISC-V\newline{}(2017) & Xilinx Artix-7\newline{}SiFive E31, U54 & multi-enclaves & changeable. \newline{}Runtime, firmware & changeable & -     & mobile-level\newline{}desktop-level & \emptycirc & confidentiality\newline{}integrity\\
    \hline
    Nitro Enclaves & AWS$^\gamma$\newline{}(2020) & Nitro Cards \& \newline{}Security Chip (EC2) & multi-enclaves & changeable. \newline{}VMs, OS/Hypervisor & up to available \newline{}instance RAM & \multicolumn{1}{p{5.585em}|}{remote\newline{}AWS service} & desktop-level\newline{}enterprise-level & \fullcirc & confidentiality\newline{}integrity\\
    \hline
    EdgeLock Secure \newline{}Enclave & NXP\newline{}(2021) & i.MX 8ULP(-CS)\newline{}i.MX 9 & \multicolumn{1}{p{6.415em}|}{single \newline{}secure world} & {small.\newline{}Firmware} & {changeable\newline{}typically small} & \multicolumn{1}{p{5.585em}|}{remote\newline{}Azure Sphere} & IoT-level\newline{}mobile-level & \emptycirc & confidentiality\newline{}integrity\\
    \hline
    Confidential Compute \newline{}Architecture (CCA) & Arm\newline{}(2022) & Armv9-A & multi-realms & changeable. \newline{}OS kernel, firmware & up to available \newline{}system RAM & remote  & mobile-level\newline{}desktop-level & \halfcirc & confidentiality\newline{}integrity\\
    \hline
    \end{tabular}%
    }
    \begin{tablenotes}
    \item[$\alpha$]{Full TCB size is implementation-specific, and the reported TCB size here is the best case supported by hardware (bare-metal potential TCB).}
    \item[$\delta$]{Most difficult for fitting ML based on prior works.} \\
    \item[$\beta$]{Strictly speaking, PMP (physical memory protection) is not a TEE but a hardware-based memory isolation feature that enables several frameworks to architect different TEEs on RISC-V.} \\
    \item[$^\gamma$]{Nitro adds new infrastructure but still builds atop existing CPUs so strictly not an ISA.}
    \item{Application Unchanged:~~\emptycirc~Major changes needed;~~\halfcirc~Light changes needed;~~\fullcirc~No change needed}
    \end{tablenotes}
\end{threeparttable}

\bigskip \bigskip
\caption{Open source TEE frameworks}\label{tab:tee_framework}
  \footnotesize
  \setlength\extrarowheight{2pt}
  \begin{threeparttable}
  \resizebox{1\textwidth}{!}{%
    \begin{tabular}{|l|l|p{11.835em}|p{9.665em}|p{16.165em}|c|p{7.835em}|c|}
    \hline
    \multicolumn{1}{|p{9.085em}|}{\textbf{Open Source \newline{}TEE Framework}} & \textbf{Leading} & \multicolumn{1}{l|}{\textbf{Supported TEEs}} & \multicolumn{1}{l|}{\textbf{On Top Of}} & \multicolumn{1}{l|}{\textbf{Main Functions}} & \multicolumn{1}{l|}{\textbf{Attestation}} & \multicolumn{1}{l|}{\textbf{Languages}} & \multicolumn{1}{p{4.585em}|}{\textbf{ML \newline{}Examples}} \bigstrut\\
    \hline
    Intel SGX SDK \cite{intelsgxsdk} & Intel & \multicolumn{1}{l|}{SGX} & \multicolumn{1}{l|}{firmware \& driver} & \multicolumn{1}{l|}{API and librarys for SGX app.} & \halfcirc & \multicolumn{1}{l|}{C/C++} & \fullcirc \bigstrut\\
    \hline
    OP-TEE \cite{optee} & Linaro & \multicolumn{1}{l|}{TrustZone} & \multicolumn{1}{l|}{firmware \& driver} & \multicolumn{1}{l|}{API and librarys for TZ app.} & \emptycirc & \multicolumn{1}{l|}{C/C++} & \fullcirc \bigstrut\\
    \hline
    \multicolumn{1}{|p{9.085em}|}{MultiZone \cite{multizone}} & \multicolumn{1}{p{9.585em}|}{HEX-five} & \multicolumn{1}{l|}{RISC-V} & \multicolumn{1}{l|}{firmware \& driver} & \multicolumn{1}{l|}{API and librarys for RISC-V Enclave app.} & \halfcirc & \multicolumn{1}{l|}{C/C++} & \emptycirc \bigstrut\\
    \hline
    \multicolumn{1}{|p{9.085em}|}{Keystone \cite{keystone}} & \multicolumn{1}{p{9.585em}|}{UC Berkeley} & \multicolumn{1}{l|}{RISC-V} & \multicolumn{1}{l|}{firmware \& driver} & \multicolumn{1}{l|}{API and librarys for RISC-V Enclave app.} & \halfcirc & \multicolumn{1}{l|}{C/C++} & \fullcirc \bigstrut\\
    \hline
    Open Enclave SDK \cite{openenclave} & Microsoft & \multicolumn{1}{l|}{SGX, TrustZone} & \multicolumn{1}{l|}{SGX SDK \& OP-TEE} & \multicolumn{1}{l|}{generlized API for TEE app.} & \halfcirc & \multicolumn{1}{l|}{C/C++} & \fullcirc \bigstrut\\
    \hline
    Asylo \cite{asylo} & Google & \multicolumn{1}{l|}{SGX} & \multicolumn{1}{l|}{SGX SDK} & \multicolumn{1}{l|}{generlized API for TEE app.} & \halfcirc & \multicolumn{1}{l|}{C/C++} & \fullcirc \bigstrut\\
    \hline
    Teaclave \cite{teaclave} & Apache & \multicolumn{1}{l|}{SGX} & \multicolumn{1}{l|}{SGX SDK} & \multicolumn{1}{l|}{API for Rust programming} & \halfcirc & \multicolumn{1}{l|}{Rust} & \fullcirc \bigstrut\\
    \hline
    Gramine \cite{gramine} & \multicolumn{1}{p{9.585em}|}{Invisible Things Lab \newline{}/ Intel} & \multicolumn{1}{l|}{SGX} & \multicolumn{1}{l|}{SGX SDK} & \multicolumn{1}{l|}{lightweight library OS to host app.} & \halfcirc & Native executable* & \fullcirc \bigstrut\\
    \hline
    Occlum \cite{occlum} & Tsinghua / Ant Group & \multicolumn{1}{l|}{SGX} & \multicolumn{1}{l|}{SGX SDK} & \multicolumn{1}{l|}{lightweight library OS to host app.} & \halfcirc & Native executable* & \fullcirc \bigstrut\\
    \hline
    \multicolumn{1}{|p{9.085em}|}{SCONE \cite{scone}} & \multicolumn{1}{p{9.585em}|}{Scontain} & \multicolumn{1}{l|}{SGX} & \multicolumn{1}{l|}{container \& driver} & \multicolumn{1}{l|}{TEE-based container image generation} & \fullcirc & Docker image* & \fullcirc \bigstrut\\
    \hline
    Enarx \cite{enarx} & Red Hat & SGX, SEV, CCA (upcoming) & \multicolumn{1}{l|}{firmware \& driver} & \multicolumn{1}{l|}{WebAssembly sandbox} & \fullcirc & Wasm binary* & \emptycirc \bigstrut\\
    \hline
    HyperEnclave \cite{jia2022hyperenclave} & Multiple inst & SGX, SEV & \multicolumn{1}{l|}{RustMonitor hypervisor} & \multicolumn{1}{l|}{cross-platform enclave} & \fullcirc & multiple & \emptycirc \bigstrut\\
    \hline
    Veracruz \cite{veracruz} & Arm   & SGX, TrustZone, Nitro \newline{}Enclaves, CCA (upcoming) & SGX SDK \& OP-TEE\newline{}Nitro Enclaves SDK & proxy attestation, policy\newline{}WebAssembly sandbox & \fullcirc & Wasm binary* & \fullcirc \bigstrut\\
    \hline
    \end{tabular}%
    }
    \begin{tablenotes}
    \item{Attestation:~~\emptycirc~Not available;~~\halfcirc~Self-configuration required;~~\fullcirc~Work off-the-shelf}
    \item[*]{Most modern languages supported by compiling as a target executable/binary.}
    \item{ML examples:~~\emptycirc~Not exist;~~\fullcirc~Exists} ~~However, One non-exist example can be developed using an existing ML framework with corresponding languages.
\end{tablenotes}
\end{threeparttable}
\end{sidewaystable*}

%% file: 4-cc-for-ml.tex
\section{Confidential Computing Solution for Machine Learning}
\label{sec:cc_for_ml}

\subsection{Overview}
\label{sec:cc_ml_overview}

\begin{figure}[t!]
    \centering
    \includegraphics[width=0.7\linewidth]{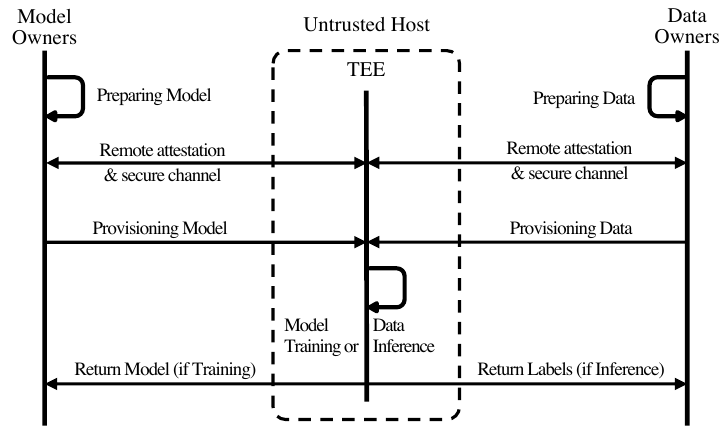}
    \caption{Overview flow of Confidential Computing which utilizes the untrusted host's Trusted Execution Environment to protect the model and data in machine learning.}
    \label{fig:overview}
\end{figure}

\begin{figure}[t!]
    \centering
    \includegraphics[width=0.7\linewidth]{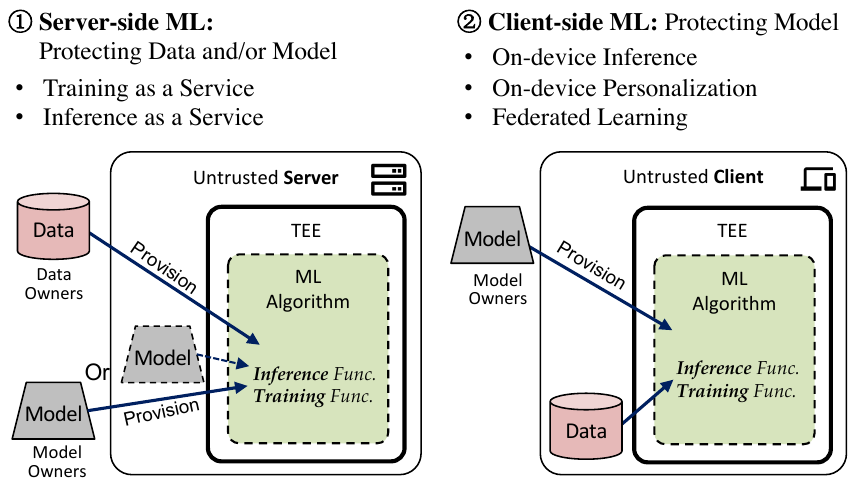}
    \caption{Server-side ML (Left) and client-side ML (Right) protection using Trusted Execution Environments. Note that, in relation to the ML paradigm defined in Section~\ref{sec:ml_paradigm}, most centralized ML is server-side, and most distributed ML is client-side, depending on how the server and the client are defined under specific ML scenarios.}
    \label{fig:protection_classes}
\end{figure}

In Confidential Computing-assisted ML, data/model owners need to secretly provision their data/model to the untrusted host's TEE (see Figure~\ref{fig:overview}). This process follows the procedure of how Confidential Computing works in general.
Specifically, after preparation of the model and/or data, the owners first perform remote attestation to assure the integrity of the remote TEE, and then, establish secure communication channels to the TEE. 
Afterward, data/models are provisioned to the TEE, where model training or inference will be performed. Then,
the produced results will be returned, \ie~a trained model will be transmitted out in training, or the data label will be returned back to the user in inference. 
Note that in practice the host, model owners, data owners, and result receivers could be distinct entities, or some may not exist depending on use cases. 
We do not consider the case that one entity is ``self-supplied'' and does not need to interact with another entity as there is no trust issue and the need for Confidential Computing.
We follow the same logic when defining the paradigm of ML (\eg~centralized and distributed ML) and further classify these cases according to the nature of the host: Server or Client (see Figure~\ref{fig:protection_classes}).

\mysub{Server-side ML protection.}
A server-like host aims to provide ML service to its customers. 
Based on the specific ML functionalities, the host can provide i)~Inference as a Service (\textbf{IaaS}), \eg~\cite{grover2018privado, lee2019occlumency, kumar2022sclera} or ii)~Training as a Service (\textbf{TaaS}), \eg~\cite{hunt2018chiron, hynes2018efficient}. 
In both cases, schemes should enable the protection of \emph{data privacy}. That is, data have to be secretly provisioned to the untrusted server's TEE no matter for inference or training. 
The private information contained in the model\footnote{Note that we do not use `model privacy' to avoid confusion, as the private information from the ML model is basically about the data used to train it, \ie~in a form of data privacy. The effort of building a model from these data is more proper to be called intellectual property.} or its \emph{intellectual property} also requires protection if it is owned by another entity (\eg~external model owners/providers), not the server. 
In this case, the model will be required to secretly provision into the TEE as well.
However, the ownership of the ML model can belong to not only external owners but also to the untrusted server itself.
The latter case will only exist for ML inference, as performing inference does not require changing the model parameters. In contrast, training leads to updates of the model parameters on data which consequently leaks private information about the data (\ie~data privacy).

\mysub{Client-side ML protection.}
A client-like host usually acts as the downstream user of a server, and it performs ML based on the ML framework/algorithm and models provided by the server.
Specifically, common use cases include i)~On-device Inference (\textbf{ODI}), \eg~\cite{bayerl2020offline}, the client conducts predictions on its own data with another individual's model that is downloaded and deployed on the device, ii)~On-device Personalization (\textbf{ODP}), \eg~\cite{mo2020darknetz}, the client personalizes the existing model based on its data for later inference, and iii)~federated learning (\textbf{IaaS}), \eg~\cite{mo2021ppfl}, the client trains the model to further contribute a global model owned by another party (\eg~server). 
In all these cases, the client owns the data; thus, this client's TEE does not need to hide these local data from itself.
However, the model is considered confidential as it could be \emph{intellectual property}, \eg~the model costs its owner or the server significant effort to train. 
More importantly, \eg~the model itself can leak private information about its previous trainers (server or other clients) as discussed in Section~\ref{subsec:attack_surface}. The TEE of one client then dedicates to providing a trusted environment in order to respect other clients' data privacy.

\subsection{Key Challenges of Adapting Confidential Computing}

Adapting Confidential Computing to ML is not trivial work. The key challenges lie in the feasibility and effectiveness of utilizing TEEs for various ML services due to TEEs' limitations.

\mysub{Constrained execution environments.}
First of all, TEEs have constrained resources for reducing the size of TCB. 
As given in Table~\ref{tab:tee_summary}, the commonly used hardware TEE, SGX, provides 128MB secure memory, while TrustZone-based TEEs, such as OPTEE or trusty, by default provide up to 16$\sim$64MiB secure memory. 
Note that the SGX 2nd~\cite{mckeen2016intel} extensions allow additional flexibility in runtime management of enclave resources (\eg~adding memory to an enclave after the enclave is built and running) and multithreaded execution within an enclave.
Further, the SGX 3rd and 4th Generation from Intel® Xeon® Scalable Processors can support up to 512GB, hence 1TB of EPC on a two-socket platform~\cite{intelProcessors, lenovosgx, intelBuildSecure}, mostly due to using Advanced Encryption Standard-XEX Tweakable Block Cipher with Ciphertext Stealing (AES-XTS) instead of previous Memory Encryption Engine (MEE), to scale EPC size. Though these CPUs are not widely available to the public, some companies are utilizing these more scalable SGX features combined with high-performant AI enhancements like Intel AMX (Advanced Matrix Extensions) to gain orders of magnitude performance boost for SGX-assisted federated training or inference.\footnote{For example, Demetics Medical Technology Co. Ltd. uses scalable SGX and oneMKL to secure its AI algorithms and IP for medical devices (oneMKL and oneDNN enhance the performance and compatibility of Intel processors and deep learning frameworks)~\cite{intelDemeticsProtects}. Another example is BigDL PPML, a solution that protects the big data and AI pipeline, created by Intel and Alibaba Cloud DataTrust~\cite{intelAlibabaBuilds}.}

Although techniques using page swapping can increase available memory, it leads to significant overhead (\eg~100$\sim$1000$\times$) when running ML~\cite{lee2019occlumency,kunkel2019tensorscone}.
Depending on the TEE software stack and architecture, the size and flexibility of protected memory vary. However, such memory sizes are still highly limited compared to the memory consumed by current ML algorithms, which can reach hundreds of MBs or GBs.

In addition, processor capabilities are limited when running in the TEE mode. For example, privilege instructions are not allowed inside SGX enclaves, which makes supporting widely used functionalities (\eg~filesystem or multi-threading) difficult or even impossible. For instance, the SGX multi-threading model relies on an untrusted OS for thread scheduling which leads the system to be vulnerable to synchronization-based attacks (\eg~AsyncShock~\cite{weichbrodt2016asyncshock}). Similarly, making the secure world over-privileged in TrustZone architecture, caused mobile vendors to only provide limited TEE services and much-restricted TEE usage.
Indeed, increasing computational resources is always preferable for ML developers but it can reduce the reliability and practicality of TEE-based solutions.

\mysub{SDK supports.}
A similar challenge arises due to the limits of trusted resources.
Modern ML framework usually requires numerous libraries and cross-compilation to support high-performant data loading and computations (\eg~matrix multiplication). However, most TEEs provide basic low-level SDKs. 
Even though open-source TEE frameworks have been developed to support sandbox/container (see Table~\ref{tab:tee_framework}), it is still hard to port all ML dependencies and necessary libraries into TEEs. For example, WebAssembly supports compiling binaries from many modern languages, \eg~Rust, which could have available ML libraries, such as autograd~\cite{rustautograd}, but they have much-limited functionality compared to Tensorflow or PyTorch~\cite{paszke2019pytorch}. 
OS-based containers can provide richer libraries, but still, the lightweight version OS cannot have a rich environment same as the normal OSs like Linux. Porting a normal OS to TEEs is also infeasible (and overkill in many use cases) because it either exceeds the TEEs' size limits or leads to a huge TCB which destroys TEEs' security benefits.

\mysub{Privacy protection effectiveness.}
Furthermore, even if one manages to deploy ML inside TEEs (for training/inference), the untrusted host itself cannot access the ML computations anymore. 
Running inside TEEs does not automatically disable all potential privacy leakages as typically the produced results will need to be passed to outside of the TEE. 
Specifically, the other parts of the ML pipeline still leak private information about the hidden process inside the TEE through various attack vectors. 
For instance, predicted data labels or trained models usually need to be transmitted out of the TEE's trust boundary to a broader area, \eg~IPC or file system, that attackers can leverage to exploit membership information or even the original data as shown in~\cite{aono2017privacy,zhu2019deep,mo2020darknetz,melis2019exploiting,oh2022deepcoffea, yang2022fsaflow}; thus, another layer of protection may still need \eg~using techniques in research~\cite{jia2019memguard,poltavtseva2023confidentiality}.
Also, one particular protection strategy could have inconsistent protection effectiveness for other ML scenarios and different types of defined privacy~\cite{mo2020layer, mo2021quantifying, mo2019towards}. 
As one example, while hiding model parameters (along with gradient updates and activations) inside TEEs can defend against DRAs, it has very low efficiency for MIAs as most membership information can leak from the model's outputs, \ie~prediction results transmitted out of the TEE~\cite{jia2019memguard, sablayrolles2019white, gu2018securing}. 
Therefore, for one specific TEE-assisted use case, the target privacy should be defined clearly, and the achieved protection efficiency also requires explicit analyses.

\subsection{Existing Solutions of Confidential ML}\label{existingsolutions}

\input{tables/tee-ml-research.tex}

Previous research has been dedicated to achieving ML with TEEs, and some aim to overcome the above challenges. We summarize previous research in Table~\ref{tab:tee_ml_research}. 
We can see the trend that much early-stage work mostly focuses on centralized ML/server-side ML until around 2020 several works started researching distributed ML/client-side ML. Most recent research is on enabling TEE functionalities on AI accelerators (\eg~GPUs, NPUs, IPUs, etc.) for secure or private ML. Note that these works on accelerators make great effort on the partitioning framework, system design, and even cryptography, in order to enable TEEs, but the main goal is to support ML; therefore, we still include them here instead of the previous section on TEE frameworks.

In addition, when there is a smaller number of mistrusting entities, guaranteeing their confidentiality/integrity can be consequently easier. 
For example, in TaaS or IaaS where only one client uses the server's TEE at one time, we need protection of only the user's properties, and therefore, can protect against all confidentiality-related attacks. In a FL case where multiple clients participate, defending against all these attacks can be very difficult because the final trained model will highly probably be passed to participating clients which consequently leaks information.

We also note that these different solutions are still measured in similar metric vectors such as system performance overload on the baseline. 
This can be execution time, memory resources, or sometimes power consumption on edge devices. The most commonly used one is execution time as a direct measure of the processor's performance. 
Nevertheless, the chosen baseline can vary a lot from native CPU execution to common TEE such as SGX, to GPU execution (as shown in Table~\ref{tab:tee_ml_research}). 
The effectiveness of the TEE's protection usually is not directly measured by standardized metrics; instead, the developers give 
``security analyses'' on their system to logically compare with native systems or other protection mechanisms.

In the following subsections, we present existing solutions with their features in detail.

\mysub{Complete ML training/inference inside TEEs.}
The most straightforward approach is to deploy a complete ML training/inference process inside TEEs. In such a case, the maximum capability of the ML task is strictly limited to the TEE's space and computing constraints. 
Much early-stage work, \eg~\cite{ohrimenko2016oblivious, hunt2018chiron}, implemented and tested on small-scale ML algorithms so probably can fit the complete inference or even training process into TEEs. 
In addition, such a full deployment is also tested in later research under FL scenarios based on the assumption that ML on clients' edge devices is usually light, while remote attestation can be further adapted to manage the agreement among all clients~\cite{quoc2021secfl}.
However, when the required secure memory of training a larger ML model exceeds the TEE's size, fitting it as a whole becomes infeasible.
Then, one strategy is to use memory space efficiently -- trading off the total number of the model's layers and the number of neurons -- and to maximize the efficiency of every ``bit'' of TEEs' secure memory for ML computation. Specifically, we list several practical approaches as follows. 
i)~Conducting inference instead of training, \eg~\cite{lee2019occlumency, gu2019yerbabuena, bayerl2020offline}. Training consumes much more computational resources because of backward propagation (\eg~memory used to save model gradients and intermediate activations).
ii)~Choosing a small batch size. A large batch size leads to large memory usage because every sample in this batch produces its own activations for all model layers.
iii)~Balancing the feature extractor (\eg~convolutional layers) and the classifier (\eg~fully connected layers)~\cite{mo2021ppfl}. A properly designed feature extractor can decrease the feature dimension but still capture key features, enabling a compact classifier to achieve good performance.

\begin{figure}[t!]
    \centering
    \includegraphics[width=0.6\columnwidth]{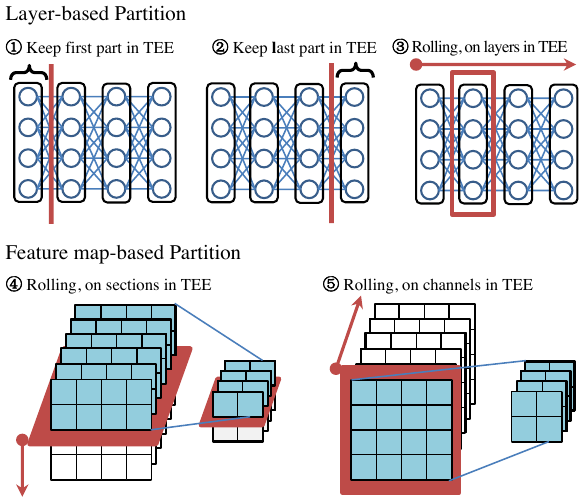}
    \caption{Partition on the ML process (shown with model architectures) in order to protect it or a part of it inside TEEs.}
    \label{fig:protection_partition}
\end{figure}

\mysub{Partitioned execution.}
To avoid exceeding the maximum secure computational resource as well as memory swapping, one approach is leveraging efficient compartmentalization (or partitioned execution) to actively optimize memory usage in the ML process.
Figure~\ref{fig:protection_partition} shows two types of partitioned execution: i)~Layer-based partition~\cite{gu2018securing, mo2020darknetz, mo2021ppfl, babar2022real, mo2022ppfl}, which works for models with layer architecture in general, and ii)~Feature map-based partition~\cite{liu2021trusted, vannostrand2019confidential, truong2021memory}, which specifically aims to convolutional layers due to the high memory cost of this type of layer.
Among all layer candidates, whether to keep the first part or the last part inside TEEs (cases \textcircled{\raisebox{-0.9pt}{1}} and~\textcircled{\raisebox{-0.9pt}{2}}) is one key choice to be confirmed, which can depend on the security/privacy protection goal. For example, protecting the first layers guarantees the privacy of \emph{original} information, while protecting the last layers guarantees the privacy of \emph{membership} information~\cite{sablayrolles2019white, aono2017privacy, mo2021quantifying}.
Rolling on layers inside TEEs (case \textcircled{\raisebox{-0.9pt}{3}}) naturally works for ML inference, because one input propagates forward throughout layers and never goes back; it works in the way that the next layers are loaded into a TEE after previous layer's computation executes finished in the TEE.
Rolling on feature maps (cases \textcircled{\raisebox{-0.9pt}{4}} and~\textcircled{\raisebox{-0.9pt}{5}}) reduce the realtime memory usage by applying \texttt{GEMM} (\ie~General matrix multiply) and \texttt{img2col} (\ie~Image to column format transform) functions on divided sections and channels of feature maps. However, this disables parallelization features and, therefore, can increase computation time.

Furthermore, although partitioned execution may be more necessary for training due to the more resources required in the backward propagation, partitioned execution for training could lead to higher overhead than inference, specifically in~\textcircled{\raisebox{-0.9pt}{1}}, ~\textcircled{\raisebox{-0.9pt}{2}}, and~\textcircled{\raisebox{-0.9pt}{3}}. Evaluations in previous work~\cite{mo2020darknetz, mo2021ppfl} indicated an increase from 10\% to 20\% approximately in their settings). The overhead is caused by a higher volume of the buffer being transmitted between the TEE and normal OS.
Note that in order to enjoy higher flexibility to support generic ML models, the partition should better not change the model architecture (unlike the model partition in the design of AlexNet~\cite{krizhevsky2012imagenet}), 

\mysub{TEE-assisted accelerators.}\label{accelerators}
In addition to optimizing the limited TEE-protected resources, one can also extend the computation power and resources to accelerators (\eg~GPUs, NPUs, TPUs, and FPGAs)~\cite{volos2018graviton,ng2021goten,zhao2022shef,xia2021sgx,park2021safe}. However, most accelerators are considered untrusted and do not support TEE-assisted functionalities. To address this issue, one way is to use CPU-TEE (\eg~SGX) to coordinate the computation delegation and to establish a secret sharing protocol between the CPU-TEE and untrusted accelerators. In such a way, accelerators are able to perform executions on a ``nonsensitive version'' of the TEE computation, \eg~masked by one-time pads~\cite{tramer2018slalom, ng2021goten}. This can also accelerate the ML's linear computation \ie~matrix multiplication. There are also TEE-assisted record-and-replay-based solutions, which are more suitable for resource-constrained and mobile GPUs. As an example, CODY~\cite{park2021safe} allows the full computation to be done inside GPU and uses TEEs like TrustZone for secure logging and inference monitoring. 
However, due to the complexity of 
heterogeneous SoC architectures, software-only secure channels between TEEs and accelerators without proper hardware root-of-trust (RoT) still leave a large attack surface and consequently weaken TEE security guarantees in turn.

Another fundamental approach is to extend accelerator hardware, systems stack (\eg~the driver), and API support for enabling accelerator with more hardware compartmentalization and TEE-assisted features. Previous works~\cite{volos2018graviton, hunt2020telekine} achieve GPU-TEE by relatively small hardware modification (unmodified GPU/CPU core but modified peripherals like PCI) and the necessary system software (\eg~runtime, hypervisor, or drivers, etc.) These solutions could add up to $41\%$ overhead on native GPUs.
Similarly, SGX-FPGA~\cite{xia2021sgx} builds a secure hardware isolation path between CPU and FPGA to protect the sensitive data in both components and in interactions. It extends the capability of the original SGX to a counterpart FPGA enclave by leveraging the FPGA's physical unclonable function (PUF) to achieve the hardware RoT on the heterogeneous SoC.
ITX (IPU Trusted Extensions)~\cite{vaswani2023confidential} follows a similar logic by extending a set of hardware and software on Graphcore Intelligence Processing Unit (IPU), a type of SOTA custom AI accelerators, with encrypted data stream through PCIe. Evaluation of the implementations showed only <5\% performance overhead on native IPUs.
This approach is also integrated into some recent NVIDIA GPUs, such as H100 Tensor Core GPU Architecture~\cite{nvidagputee}. 

More specifically, NVIDIA introduced Multi-Instance GPUs (MIG) like Hopper~\cite{andersch2022nvidia}, which enables each GPU to be partitioned into several smaller, fully isolated instances with their own memory, cache, and compute cores. This HW-assisted compartmentalization approach allows more flexible implementation of trusted computing domains with trusted IO paths. The Hopper architecture further enhances MIG by supporting multi-tenant, multi-user virtualized domains across up to seven GPU instances, securely isolating each instance with confidential Computing at the hardware and hypervisor level. For instance, it provides dedicated video decoders for each MIG instance to deliver secure, high-throughput intelligent video analytics (IVA) on shared infrastructure, as well as concurrent MIG profiling, for secure optimization of resource allocation.

Hence, despite the current limitations, Confidential Computing with more flexible features would be supported on future accelerators. At the same time, ML frameworks and systems must still be prepared to utilize these features properly, which can not be merely resolved from the userspace perspective in an efficient and high-performant way due to the limitations of current systems abstractions that are not designed with such hardware-security features in mind. For instance, previous work shows how extending the underlying OS and system stacks for enabling more extensible and TEE-aware compartmentalization, called dispersed compartments~\cite{tarkhani2022secure}, could facilitate properly utilizing such hardware-based compartmentalization features.

\mysub{Attack-based privacy measurement.}
The TEE protection needs measurements to show its effectiveness. Currently, the main approach for measuring TEE security is through red-teaming and performing various attack vectors directly. Whether to and which attack to perform mostly depends on the threat model defined for the TEE which could be inadequate for different ML use cases. As one example, regarding the original input protection with TEEs, one performs attacks to reconstruct input data~\cite{gu2018securing} while there is no need to conduct membership inference attacks. By contrast, TEE protection on membership information does not take care of other private information and only be measured with membership inference attacks~\cite{mo2020darknetz}. However, due to the poor definition of various types of privacy, the attack-based analysis lacks a theoretical foundation on privacy leakage which sequentially makes the TEE-based privacy guarantee less trustworthy.

%% file: tables/tee-ml-research.tex
\begin{table*}[!tp]
\caption{Previous research that uses Trusted Execution Environments to guarantee the confidentiality/integrity of machine learning.}\label{tab:tee_ml_research}%
  \centering
  \footnotesize
  \setlength\extrarowheight{1pt}
  \begin{threeparttable}
  \resizebox{1\textwidth}{!}{
    \begin{tabular}{|p{8.75em}|l|l|l|c|c|c|l|p{16.585em}|}
    \hline
    \multirow{2}[4]{*}{\textbf{Work \& Year}} & \multicolumn{2}{c|}{\textbf{Confid. Computing}} & \multicolumn{5}{c|}{\textbf{Machine Learning}} & \multicolumn{1}{c|}{\multirow{2}[4]{*}{\textbf{Advanced Features}}} \bigstrut\\
\cline{2-8}    \multicolumn{1}{|c|}{} & \multicolumn{1}{c|}{\textbf{TEE Type }} & \multicolumn{1}{p{8em}|}{\textbf{SDK/Partitioning}\newline{}\textbf{Framework}} & \multicolumn{1}{p{4.25em}|}{\textbf{DL \newline{}Library}} & \multicolumn{1}{p{2.835em}|}{\textbf{Train\newline{}-ing}} & \multicolumn{1}{p{8em}|}{\textbf{Paradigm, Loc., \&}\newline{}\textbf{No. of entities}$^\alpha$} & \multicolumn{1}{p{2.915em}|}{\textbf{Protect\newline{}Against}} & \multicolumn{1}{p{7.335em}|}{\boldmath{}\textbf{Overhead (execution time)$^\beta$}\unboldmath{}} & \multicolumn{1}{c|}{} \bigstrut\\
    \hline
    Obliv. MP'16 \cite{ohrimenko2016oblivious} & SGX   & Intel SGX SDK & \multicolumn{1}{p{4.25em}|}{fast CNN} & \fullcirc & TaaS, Server, Multiple     & C(DR, AI)  & $1-3$\% on Native & \multicolumn{1}{l|}{Data-obliviousness} \bigstrut\\
    \hline
    Chiron'18 \cite{hunt2018chiron} & SGX   & Intel SGX SDK & Theano  & \fullcirc & TaaS, Server, Single     & C(All), I(MP)   & $4-20$\% on Native & Multi-enclaves, Ryoan VM \bigstrut\\
    \hline
    PRIVADO'18 \cite{grover2018privado} & SGX   & Intel SGX SDK & \multicolumn{1}{p{4.25em}|}{ONNX} & \emptycirc & IaaS, Server, Multiple   & C(DR, AI)    & $\sim17.18$\%  on Native & Compiler, data-obliv. \bigstrut\\
    \hline
    Myelin'18 \cite{hynes2018efficient} & SGX   & Intel SGX SDK & TVM   & \fullcirc & TaaS, Server, Multiple   & C(DR, AI)   & $4.7-9.2$\% on Native & DP for data obliviousness \bigstrut\\
    \hline
    Slalom'18 \cite{tramer2018slalom} & SGX   & Intel SGX SDK & Eigen & \emptycirc & IaaS, Server, Single   &  C(All), I(MP) & SU. $4-20\times$ on SGX & Freivald’s algorithm + GPU \bigstrut\\
    \hline
    Graviton'18 \cite{volos2018graviton} & GPU-TEE & \multicolumn{1}{p{6.835em}|}{CUDA RT driver} & Caffe & \fullcirc & TaaS, Server, Single   &  C(All), I(MP)   & $17-33$\% on GPUs & Secure CUDA for TEE on GPUs \bigstrut\\
    \hline
    Occlumency'19 \cite{lee2019occlumency} & SGX   & Intel SGX SDK & Caffe & \emptycirc & IaaS, Server, Single  &  C(All), I(MP)  & $\sim72$\%  on Native & On-demand loading, Channel partition \bigstrut\\
    \hline
    TensorSCONE \cite{kunkel2019tensorscone} & SGX   & Intel SGX SDK & TF    & \fullcirc & TaaS, Server, Single   & C(All), I(MP)   & $\sim3\times$ on Native & Docker supported, Compiler \bigstrut\\
    \hline
    YerbaBuena'19 \cite{gu2019yerbabuena} & SGX   & Intel SGX SDK & Darknet & \emptycirc & IaaS, Server, Single     & C(DR)     & up to 7.5\% on Native & Layer-wise partition, Privacy measure \bigstrut\\
    \hline
    Origami'19 \cite{narra2019privacy} & SGX   & Intel SGX SDK & Eigen & \emptycirc & IaaS, Server, Single     & C(DR)     & SU. $2-15\times$ on CPUs & Model partitioning, GAN attacks \bigstrut\\
    \hline
    OMG'20 \cite{bayerl2020offline} & TrustZone & SANCTUARY & \multicolumn{1}{p{4.25em}|}{TFLM} & \emptycirc & ODI, Client, Single   & C(MS)   & $2.11\times$ on Native & \multicolumn{1}{l|}{-} \bigstrut\\
    \hline
    DarkneTZ'20 \cite{mo2020darknetz} & TrustZone & OP-TEE & Darknet & \fullcirc & ODP, Client, Single    & C(MI)     & $3-10$\% on Native & Layer-wise partition, Privacy measure \bigstrut\\
    \hline
    \multicolumn{1}{|l|}{TrustFL'20 \cite{zhang2020enabling}} & SGX   & Intel SGX SDK  & TF    & \fullcirc &   ODP, Client, Single    &  I(MP)    & $\sim2\times$ on Native & GPU-outsourcing, Random sampling \bigstrut\\
    \hline
    Telekine'20 \cite{hunt2020telekine} & GPU-TEE & \multicolumn{1}{p{6.835em}|}{ROCm \& CUDA} & MXNet  & \fullcirc & TaaS, Server, Single     & C(All), I(MP)   & $10-41$\% on GPUs & Timing attacks, data-obliv. \bigstrut\\
    \hline
    Serdab'20 \cite{elgamal2020serdab} & SGX   & Intel SGX SDK & TFlite & \emptycirc & IaaS/ODI, S+C, Single  &  C(All), I(MP)  & SU. $4.7\times$ on one SGX & Muliple TEEs, Resource management  \bigstrut\\
    \hline
    HybirdTEE'20 \cite{gangal2020hybridtee} & SGX + TZ   & OP-TEE + SIGMA & Darknet & \emptycirc & IaaS/ODI, S+C, Single  &  C(All), I(MP)  & SU. $1-6\times$ on Client & Hybrid TEEs, Partitioned ML \bigstrut\\
    \hline
    AegisDNN'21 \cite{xiang2021aegisdnn} & SGX   & Intel SGX SDK & TF, PyTorch & \emptycirc & IaaS, Server, Single  &  I(MP)  & Realtime service & Safety Profiling of Layers  \bigstrut\\
    \hline
    MLCapsule'21 \cite{hanzlik2021mlcapsule} & SGX   & Intel SGX SDK & Eigen  & \emptycirc & ODI, Client, Single   &  C(MS)   & $55-116$\% on Native & \multicolumn{1}{l|}{-} \bigstrut\\
    \hline
    Trusted-NN'21 \cite{liu2021trusted} & TrustZone & OP-TEE & \textit{Not reported}   & \emptycirc &  ODI, Client, Single  & C(MS)   & $22.8$\% on Native & Weights \& Feature-map partition \bigstrut\\
    \hline
    Mem-Eff.'21 \cite{truong2021memory} & SGX   & \multicolumn{1}{p{6.835em}|}{Azure CC (VM)} & Darknet & \emptycirc & ODI, Client, Single   & C(MS)   & $9-100$\% on Native & Channel \& Y-plane partition \bigstrut\\
    \hline
    PPFL'21 \cite{mo2021ppfl} & \multicolumn{1}{p{4.415em}|}{SGX + TZ} & \multicolumn{1}{p{6.835em}|}{OE + OP-TEE} & Darknet & \fullcirc & FL, S+C, Multiple & C(All)   & $\sim15$\% on Native & Layerwise training, Privacy measure \bigstrut\\
    \hline
    SecFL'21 \cite{quoc2021secfl} & \multicolumn{1}{p{4.415em}|}{SGX} & \multicolumn{1}{p{6.835em}|}{SCONE} & TF & \fullcirc & FL, Client, Multiple  & C(All)   & \textit{Not reported} & Remote attestation \bigstrut\\
    \hline
    Goten'21 \cite{ng2021goten} & SGX & Intel SGX SDK  & Eigen & \fullcirc & TaaS, Server, Single   & C(All), I(MP)   & SU. $6.84\times$ on SGX & GPU-outsourcing, Non-colluding servers \bigstrut\\
    \hline
    Citadel'21\cite{zhang2021citadel} & SGX   & \multicolumn{1}{p{6.835em}|}{SCONE} & \multicolumn{1}{p{4.25em}|}{TF} & \fullcirc & TaaS, Server, Multiple    &  C(DR,AI,MS)  & $9-73$\% on Native & Multi enclaves for training \bigstrut\\
    \hline
    SecDeep'21 \cite{liu2021secdeep} & TrustZone & ATF \& OP-TEE & ARM NN DL & \emptycirc & ODI, Client, Single & C(MS) & 16$\times$ - 172$\times$ on Native & GPUs, Controlling kernel page table  \bigstrut\\
    \hline
    Fair. Audit'22 \cite{park2022fairness} & SGX  & SGX-LKL-OE & PyTorch & \emptycirc & IaaS, Server, Single    & C(All), I(MP) & $\sim30$\% on Native & Fairness audit, Modler + regulator \bigstrut\\
    \hline
    GuardNN'22 \cite{hua2022guardnn} &  FPGA-based  & (Sim.) CHaiDNN &  SCALE-Sim & \fullcirc &  TaaS, Server, Single   & C(All), I(MP) & up to 1.07\% on Native & - \bigstrut\\ 
    \hline
    SOTER'22 \cite{shen2022soter} &  SGX  & Gramine & PyTorch & \emptycirc & ODI, Client, Single & C(MS), I(MP) & up to 1.27$\times$ on GPUs & Oblivious fingerprinting for integrity \bigstrut\\ 
    \hline
    StrongBox'22 \cite{deng2022strongbox} & TrustZone & Midgard \& ATF & ARM NN DL & \emptycirc & ODI, Client, Single & C(MS) & 2.8\% - 19.7\% on GPUs & Fine-grained memory protection policy \bigstrut\\ 
    \hline
    Dash'23 \cite{sander2023dash} & SGX & Intel SGX SDK & ONNX & \emptycirc & IaaS, Server, Single & C(All), I(MP) & SU. 5 - 140$\times$ on CPUs & Garbled circuits, MPC, GPUs \bigstrut\\
    \hline
    ITX'23 \cite{vaswani2023confidential} & IPU-TEE & IPU/CCU Stack & TF & \fullcirc & TaaS, Server, Single  & C(All), I(MP) & <5\% on IPUs & remote attestation, AI accelerator \bigstrut\\ 
    \hline
    Honeycomb'23 \cite{mai2023honeycomb} & SEV-SNP & SVSM \& SM \& HIP & PyTorch, etc. & \fullcirc & TaaS, Server, Single  & C(All), I(MP) & $\sim2\%$ on GPUs & Software-based, Small TCB \bigstrut\\ 
    \hline
    T-Slices'23 \cite{islam2023confidential} & TrustZone & OP-TEE & Darknet & \emptycirc & ODI, Client, Single  & C(MS) & SU. $29\%$ on DarknetTZ & Dynamic fragmentation of DNNs \bigstrut\\ 
    \hline
    AvaGPU'23 \cite{wang2023secure} & TrustZone & OP-TEE + CUDA & TensorRT & \emptycirc & ODI, Client, Single  & C(MS), I(MP) & $\sim15.87\%$ on GPUs & real-time CPS system, Small TCB \bigstrut\\ 
    \hline
    SecureLoop'23 \cite{lee2023secureloop} & \multicolumn{3}{|c|}{TEE-like DNN Accelerators in general} & \fullcirc & \textit{N/A}  & \textit{N/A} & SU. $\sim33.2\%$ on GPUs & Design space, Cryptographic engines \bigstrut\\ 
    \hline
    Deluminator'23\cite{tarkhani2023information} & \multicolumn{3}{|c|}{SGX \& TrustZone} & \emptycirc & ODI, Client, Single  & 
    C(All), I(MP) & $\sim7\%-29\%$ & Information flow tracking, Auditing \bigstrut\\ 
    \hline

    PoUL'24 \cite{weng2024proof} & SGX & Intel SGX SDK & Intel DNNL & \fullcirc & TaaS, Server, Single  & C(All), I(MP) & \textit{N/A} & Unlearning \& Proof of unlearning \bigstrut\\ 
    \hline
    CAGE'24 \cite{wangcage} & CCA & RMM \& SM & \textit{Not reported} & \emptycirc & IaaS/ODI, S/C, Single  & C(All), I(MP) & $1.2\% \sim7.6\%$ on GPUs & Unified-memory GPUs, Small TCB \bigstrut\\ 
    \hline
    \end{tabular}%
    }
    \begin{tablenotes}
    \item[$^\alpha$]{Paradigm: Categories following Figure~\ref{fig:protection_classes}; Loc.: Locations of used TEEs; No. of entities: Number of mistrusting participating entities to be protected.} \\
    \item[$^\beta$]{The reported overhead is given to show general performance and \emph{cannot} be cross-compared due to measuring under different experimental settings.} \\
    \item[SU.]{refers to speed-up, instead of overhead.}
    \item{Training:~~\emptycirc~Not available;~~\fullcirc~Available} \\
    \item{Protect Against:~~C(DR) = Confidentiality - Data Reconstruction;~~C(AI) = Confidentiality - Attribute Inference;~~C(MI) = Confidentiality - Membership Inference;}\\
    \item{\hspace{1.65cm} C(MS) = Confidentiality - Model Stealing, I(MP) = Integrity - Model Poisoning; ~~I(DP) = Integrity - Data Poisoning}\\
    \item{Note that the difficulties of achieving one protection aim also depend on the number of mistrusting participated entities to be protected; Here we refer \\ 
    to the state that all participating entities' confidentiality/integrity are guaranteed if filled.}
\end{tablenotes}
\end{threeparttable}
\vspace{-5pt}
\end{table*}%

%% file: 5-cc-for-ml-integrity.tex
\section{Confidential Computing Helping Machine Learning Integrity}
\label{sec:cc_for_ml_integrity}

Intuitively, other than privacy guarantee, deploying ML training/inference process into TEEs could avoid malicious modifications to this process and therefore ensure integrity.
However, in practice, as the other parts of the ML pipeline still can be breached, protecting only the training/inference stage does not always guarantee integrity. For instance, the training happening outside can be modified maliciously in partitioned execution while tackling space-constrained problems. 
In this section, we present the key challenges in enabling ML integrity using TEEs and the solutions that exist or can be adapted.

\subsection{Key Challenges in Enabling ML Integrity}

\mysub{Large attack surface.}
There is a large attack surface to breach integrity; however, one cannot expect to deploy the complete ML lifecycle inside TEEs to achieve an integrity guarantee due to the TEEs' resource constraints.
These limitations for achieving a proper integrity guarantee are even more severe compared with the issue we met when guaranteeing confidentiality. 
Specifically, while an honest-but-curious (\ie~confidentiality-related) adversary has the goal to explore specific private information, the integrity-related adversary can be more diverse -- it aims at not only actively changing the ML process but also even breaking the process by covertly changing one bit in updated weights/gradients. Indeed, to disclose confidential information, adversaries usually explore the target data/model, and contained information in the model may also gradually decrease along with more aggregated information in training.
However, to breach integrity, one just needs to change one step in the training process (even in one bit) among participants (\eg~Byzantine attacks~\cite{fang2020local} or synchronization attacks~\cite{sanchez2020game}). This malicious change can also be triggered at other stages \eg~data preparation.

\mysub{Uncontrolled input/output space.}
While deploying ML training/inference inside TEEs avoids unauthorized direct changes in the produced result/models, the upstream pipeline, \ie~data preparation, or input space, is uncontrolled. Editing the inputs and their labels can affect the later ML process (\eg~called ``dirty'' inputs/labels), which is also the practice of performing some attacks, \eg~poisoning attacks or adversarial example attacks.
Adversarial examples~\cite{goodfellow2014explaining, moosavi2017universal} that add calibrated noises to the input can seriously perturb the integrity of the input space due to the hardness of detecting the noisy perturbation.
In addition to that, the downstream pipeline can be edited maliciously. For example, after the ML algorithm inside the TEE makes a prediction and transmits this result out, a server-side adversary fakes the result, which is also hard to detect on one result receiver side.

\mysub{Low interpretability of ML.}
One major reason for the difficulty of detecting integrity perturbations is the low interpretability of current ML/DL~\cite{li2021interpretable}. Specifically, current practitioners train ML models using SGD, \ie~weights of neurons are updated iteratively to find the optimal solution. There is still no comprehensive theoretical framework to fully interpret the internal steps of ML training. Therefore, one cannot determine whether the weight value of a neuron is uncompromised or not without \emph{re-running} the ML process using the same setting.
With such low interpretability, malicious changes such as the adversarial perturbation in inputs and the changes in model training are hard to trace.
Algorithm-based approaches have been proposed to alleviate this situation \eg~by removing outliers that are considered as unreasonable bits~\cite{fang2020local}, by further training the model on adversarial inputs~\cite{tramer2018ensemble}, or by cryptography protocols~\cite{chowdhury2021eiffel}. However, utilizing a system-based isolation technique (\ie~TEEs) to deal with the interpretability issue is not straightforward. One may still need to find a way to identify and ensure key elements of the ML pipeline are trustworthy using TEEs, \ie~rerunning and verifying sensitive parts of the ML.

\subsection{Existing/Adoptable Solutions}

\mysub{Assurance mechanisms.}
To assure the integrity of ML training/inference, one way is to rerun the complete or a part of the process inside TEEs.
For example, the heaviest computation in ML is the matrix multiplication which is preferable to be outsourced to distrusted GPUs for acceleration. To verify whether this matrix multiplication is honestly performed, \cite{tramer2018slalom}'s work use Freivald's algorithm~\cite{motwani1996randomized} which utilizes randomization to reduce time complexity from $\mathcal{O}(n^{2.373})$ (best-known matrix multiplication algorithm) to $\mathcal{O}(kn^2)$ with a probability of failure less than $2^{-k}$.
Besides, based on the nature of iterative training of ML, one can also sample and verify only a fraction of the training process in TEEs to reduce the overhead~\cite{zhang2020enabling, zhang2022adversarial}. This will require backups for every several training iterations to reach a checkpoint faster, where Merkle hash tree-based method~\cite{merkle1980protocols} can be used to reduce the storage overhead. 
With such an assurance scheme, reference~\cite{zhang2020enabling} demonstrates that adversaries only have a 1\% probability of successfully cheating even despite that they honestly completed 90\% of training rounds. 
Furthermore, there could exist more efficient ways of sampling or verification based on watermarking~\cite{lukas2022sok}, but how such an assurance guarantees the integrity of other stages in the ML pipeline deserves more investigation.

\mysub{Input/output space control.}
The integrity of input/output in the ML pipeline needs to be assured considering that the input/output space is far more uncontrollable than the training/inference process. 
One way that is generally used for data assurance and can also be adopted in ML input control is to require a digital signature for underlying data generation. This can be done by hashing the generated sensor data (inside TEEs for example) to produce hash values. After that, sensor data are authenticated using these hash values~\cite{merkle1989certified, cohen1987cryptographic,karapanos2016verena}. At a later stage, the digital signature would also work in a similar way for remote attestation when outputting prediction results, considering that the result receiver trusts the TEE's behavior.
Nevertheless, it becomes harder especially for supervised learning because the data generation usually involves human annotations that are out of the control of the digital signature.

Even in semi-supervised/unsupervised learning, it is possible to generate adversarial data without breaching sensors/TEEs, \eg~by changing physical surroundings or doing abnormal behaviors that the ML does not expect to learn. Therefore, the exploration of how to assure the integrity of the input space using TEEs is still a research question. One example can be adopting detection mechanisms (\eg~\cite{brodley1999identifying, chu2016data, bahri2020deep} to identify fake/dirty inputs inside TEEs, which deserves further investigation.
Previous work also introduced system-wide TEE-assisted information flow control, particularly over peripherals and OS services to detect and mitigate such unauthorized IO-based attack vectors~\cite{tarkhani2020enclave}. However, these techniques require fundamental changes in system software including commodity OSs.

%% file: 6-tee-limitation.tex
\section{Limitations of existing TEE systems}
\label{sec:limit_tee}

Despite the benefits of TEEs/enclaves, the right abstractions for securely and efficiently utilizing them are still not clear. Particularly, when combining various TEE solutions (\eg~for multi-platform use cases) or integrating with existing systems. Below we further detail the limitations of existing TEE systems that obstruct their use in ML.

\begin{figure*}[t!]
\centering
\hspace{-1cm}
    \includegraphics[width=1\linewidth]{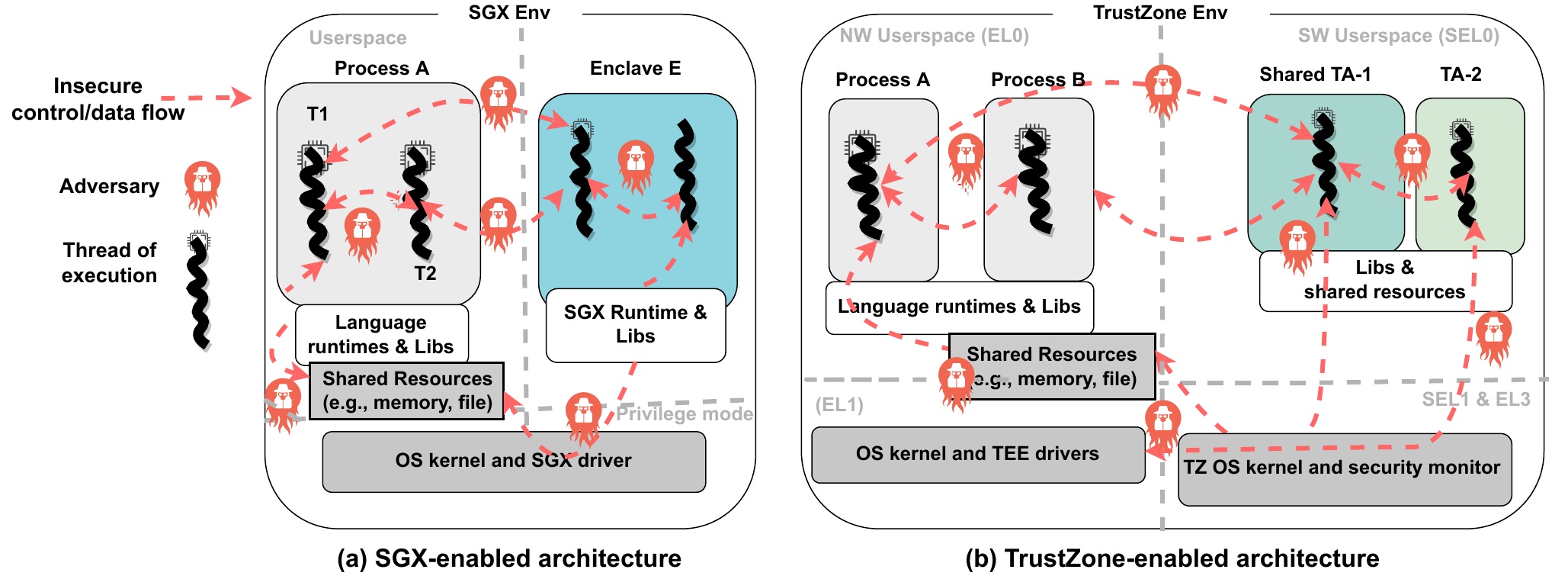}
  \caption{Unsafe or malicious data/control flows in TEE systems cause various attack vectors within and across trust boundaries in TEE systems. An adversary could exploit vulnerabilities within almost all layers of abstractions and software stacks, as shown in previous research. This figure simplifies some of these unsafe control and data flows for launching an attack or propagating different vulnerabilities to compromise the confidentiality or integrity guarantees of a TZ-assisted or SGX-assisted system.}
\label{fig:tzthreats}
\vspace{-5pt}
\end{figure*}

\subsection{Widening the Attack Surface}

\mysub{New features lead to new attack vectors.}
TEEs require rich functionalities and performance features to support running modern ML programs. However, merely adding hardware features without a comprehensive security model and mechanism with existing system software and privilege boundaries (\eg~sharing data or resources) can even increase the system's complexity and open new attack vectors. Particularly in TEE architectures which the untrusted host OS or hypervisor is responsible for managing TEE/enclave system resources. For instance, recent work such as SGXRacer~\cite{Chen:2023:SGXRacer} shows more than 1000 new vulnerabilities through controlling/misusing kernel scheduler and SGX shared (\ie~the racing) variables.
Also, recent security evaluations of Intel TDX have revealed various vulnerabilities in different layers of the TDX software stack\cite{Intel_TDX_2023}, including its attestation services; which also include SGX-based quoting enclave, during TD live migration (a new feature) and TD binding\cite{Intel_SA_01036} which may allow information leak and escalation of privilege (e.g., see CVE-2023-45745 and CVE-2023-47855).

As another example, ARM TrustZone introduced hardware support for secure and non-secure worlds by adding several privilege levels in each world. However, the architecture limitations, such as the insecure sharing mechanism and fixed/one-way security model~\cite{tarkhani2019snape}, combined with insecure software designs lead to new attack vectors, particularly when sharing resources or exchanging data/control between different privilege layers~\cite{machiry2017boomerang,cerdeira2020sok}. 
Figure~\ref{fig:tzthreats}.b demonstrates some of these attack vectors in a TrustZone-assisted environment. 

Previous attacks, such as Boomerang~\cite{machiry2017boomerang} and HPE (Horizontal Privilege Escalation)~\cite{suciu2020horizontal}, also show how the semantic gap between heterogeneous compartments (\eg~userspace processes and TrustZone TAs (Trusted Apps)) leads to severe privilege escalation threats. Boomerang exploits a confused deputy vulnerability inside a TA via the shared memory between the two worlds and takes advantage of secure world privilege to make the host kernel memory accessible. Figure~\ref{fig:tzthreats}.b shows simplified data/control flows of such attacks through compromised user space processes or libraries to a vulnerable TA via shared resources/RPCs and then through that compromised TA to the secure kernel for finally taking control of the system.
Similarly, HPE attacks demonstrate that an adversary process could compromise another userspace process through a shared TA (e.g., a shared crypto TA) or insecure privilege management between two TAs. For example in Figure~\ref{fig:tzthreats}.b an adversary could misuse the $TA_1$'s shared persistent state and logs between process $A$ and $B$ to leak sensitive information from one process to another or propagate through $TA_2$ or other unauthorized sources.

\mysub{Huge TCB caused by in-enclave LibOS.}
Following the in-enclave LibOS approach for porting unmodified, yet complex applications into enclaves, results in a huge TCB through porting all dependencies inside the enclave (and more recently inside the entire virtual machine (VM) via TDX or SEV features). 
Such designs force developers to run large code in a single enclave, resulting in inefficient and over-privileged enclaves. This leads to wide range of new attacks due to exploiting in-enclave vulnerabilities~\cite{weichbrodt2016asyncshock,khandaker2020coin}, insecure interactions with outside~\cite{checkoway2013iago,van2019tale,khandaker2020coin}, or misusing enclaves to escalate privileges~\cite{machiry2017boomerang,suciu2020horizontal,weiser2019sgxjail}. Currently, there is no systematic way to neither detect nor protect against these threats.

\mysub{Vulnerable interactions with outside.}
None of the existing partitioning frameworks fully consider the complex attack surface originating from insecure interactions between the host hypervisor, OS, userspace processes, and enclaves. 
For instance, an untrusted/malicious hypervisor can inject virtual interrupt to affect AMD SEV-SNP and AMD SEV-ES (CVE-2024-25742), compromise guest memory integrity through improper address validation or memory safety issues
(CVE-2023-20566 and CVE-2023-20519).
As another example, Vicarte \etal~\cite{sanchez2020game} show how asynchronous poisoning attacks on TEE systems lead to changing the accuracy and integrity violation of in-enclave ML models.
Civet uses dynamic taint-tracking to control the flow of objects on enclave interfaces, but cannot help for proper privilege separation and against more complex attacks (\eg~HPE attacks~\cite{suciu2020horizontal}) and other languages.
Moreover, they do not consider in-address space compartmentalization as the key issue of handling over-privileged enclaves. 
Sirius~\cite{tarkhani2020enclave} shows that targeting these attacks requires fundamental changes to underlying OS and TEE systems. Note that developers sacrifice performance and rich functionality for the security benefits of TEEs; which could be a huge cost without actual benefit considering the insecure designs of existing TEE systems. 

\mysub{ML inside TEEs but still being attacked.}
Existing vulnerabilities of TEEs still leave the deployment of ML under some forms of attacks. For example, by observing the access pattern to TEEs, previous work in Privado~\cite{grover2018privado} can disclose the target class information \ie~classifying encrypted inputs with high accuracy (97\% and 71\% for MNIST and CIFAR10 respectively). Also, side-channel attacks on TEEs like timing, power, and Electromagnetic analysis could compromise the ML model privacy on some levels, \eg~reconstructing the model architecture and even model parameters~\cite{liu2020ganred, xiang2020open, batina2019csi}. Some of such attacks can be out of scope because our threat model does not include side-channel attacks. However, this still reflects the current limitations of the existing TEE system which deserve further investigations in future design (\eg~efficient oblivious operations for ML~\cite{ohrimenko2016oblivious, zheng2017opaque}).

\subsection{Programmability \& Performance}
The quality of ML solutions highly depends on the proper processing of 
an enormous amount of data. However, almost all TEE systems are not designed considering this essential requirement. Many TEEs lack minimal hardware resources for efficient ML programming on modern CPUs as we explained earlier (Section~\ref{sec:cc_for_ml}).
Currently, resolving these limitations causes significant additional performance overhead. As an example, MPTEE~\cite{zhao2020mptee} proposes a mechanism for supporting more flexible memory management and dynamic permission enforcement inside enclave memory. It uses MPX's three bound registers to offer the six common memory permissions (\ie~RWX, RW, RX, R, X, non-permission) with about 35\% overhead on average and 57\% in some cases. Besides, Occlum~\cite{shen2020occlum} proposed an MPX-based SFI technique for supporting in-enclave multi-processing, which adds a 10$\times$ slowdown compared to Linux processes. SGXv2 adds special instructions (\eg~\texttt{EMODPR} and \texttt{EMODPE}), to facilitate some of these issues including dynamic memory management. However, the overhead is not clear yet due to the lack of hardware support, and SGXv1 is still the dominating version for a while. 

These improvements are not sufficient for most ML use cases. First, current designs cannot be integrated with a wide range of accelerators such as GPUs, FPGAs, and TPUs. Despite recent efforts for supporting TEEs in these accelerators as described in Section~\ref{existingsolutions}, their systems software and ML framework designs and evaluations are in the early phases~\cite{jang2019heterogeneous,volos2018graviton,xia2021sgx,hunt2020telekine}. \cite{akram2022sok}'s SoK paper also shows why the existing TEEs are unsuitable for high-performance computing systems; even for new TEE architecture like CCA, improvements are still required to facilitate its use in protecting the entire ML deployment pipeline~\cite{siby2024guarantee}.
Depending on the level of protection in different ML stages, the overhead and toolchain supports can be highly diverse. Also, none of the previous work offers scalable solutions with multiple accelerators, which deserve further investigation.

It is worth noting that (which is usually ignored), programming ML solutions with TEE systems has a high learning curve due to the huge differences between their architectures and programming logic. The complexity of the integration requires expertise in both fields that may need years to build for developers. Although proficient developers can build preliminary solutions, due to the fast changes in ML software stacks, it is still hard to maintain the pipeline; thus the solution cannot be scaled to a larger community.

\subsection{Heterogeneity \& Migration}
Heterogeneous SoC architectures enable a wide range of functionalities and are now available even for modern IoT/edge platforms~\cite{zhao2022vsgx}. Modern SoCs contain heterogeneous CPUs (\eg~a combination of ARM and RISC-V architectures) and peripherals. As a result, the system stack on such devices includes multiple OSs (\eg~Linux and FreeRTOS), hypervisors, and different TEEs. As a simple example, recent NXP's i.MX boards, simultaneously support different TrustZone implementations for mixed cortex-A and cortex-M cores (\eg~i.MX 8ULP or 8QuadXPlus). Secure hardware partitioning and sharing hardware resources within such ever-growing complexity is challenging. That is the reason why most systems software (\eg~Android~\cite{androidimx}) does not expose heterogeneous TEEs to userspace applications despite hardware availability.

Moreover, heterogeneous TEEs could have different security models and resource constraints which make the migration of TEE-based ML solutions from one platform to another much more challenging. For instance, remote attestation supports in TrustZone-based and SGX-based solutions are different as well as their system stacks (including the support for fuzzing, debugging, auditing, etc.). Naively, combining different security protocols does not lead to a secure solution. Also, as summarized in Figure~\ref{fig:tzthreats}, vulnerabilities can easily be introduced and propagated in various privilege boundaries (within the different or same address spaces). Detecting and protecting against such a complex attack surface is one of the primary obstacles to securely handling TEE migrations. Ad-hoc approaches are neither practical nor scale well. We need a principled approach to deal with heterogeneous TEEs efficiently and consider future security threats~\cite{tarkhani2023information}. A naive migration could widen or worsen the attack surface while reducing the performance of ML solutions in most cases~\cite{tarkhani2022secure}.

%% file: 7-beyond.tex
\section{Next generation and Beyond}
\label{sec:beyond}

In this section, we present the potential approaches to tackle above discussed issues, toward better privacy-preserving machine learning atop TEEs.

\subsection{Strengthen privacy foundation in Confidential Computing}
\label{sec:privacyfoundation}
As one TEE's primary protection goals, privacy needs to be strictly defined and preferably be defined with theoretical basics (\eg~information theory). 
While most privacy measures are empirically evaluated by conducting attacks\footnote{Measuring privacy using attack success rates fails to explain and quantify privacy leakages, and then leads to ``Arms Race'' where current protections need re-evaluation or re-design when new attacks happen.}, defining privacy theoretically is complicated due to the broad existence of attacks in ML; one rigorous definition can fail to explain other attacks.

Among candidates, \emph{differential privacy} is a well-accepted definition of privacy. 
DP has been massively explored when applied to ML. 
Privacy audits using DP can obtain a lower bound but usually require running the training/estimation algorithm hundreds of times. One promising work~\cite{steinke2023privacy} explores the parallelism of adding or removing multiple training examples independently and reduces the privacy auditing into one training run.
However, it still faces the same generalization problem.
As one example, one can apply DP-training~\cite{abadi2016deep} to show the protection efficiency against DRAs that recover original training data from gradients computed on it. As expected, this protection is not reasonable since typical DP-training guarantees the ``membership'' privacy of the trained final model, not (and far from) intermediate gradients. 
To measure on the pre-gradient level, one needs to adjust the DP's granularity from sample-level (\ie~sensitivity of samples in the dataset) to feature-level (\ie~sensitivity of features in the sample). This level of DP noises will highly likely destroy the model utility but potentially provide a theoretical-based sense of the level of privacy for such protection. 

Another line of work is to apply computational constraint-based $\mathcal{V}$ \emph{usable information} theory~\cite{xu2019theory}. $\mathcal{V}$-information allows us to consider an attack family with bound computational power and limited knowledge. With the defined attack family, one can measure the capability of the complete attack family, which may include many attacks, by computing $\mathcal{V}$-information similar to Shannon mutual information. This $\mathcal{V}$-information then reflects the privacy leakage~\cite{mo2021quantifying}. Although such a method provides a theoretical base for measuring privacy, how to define a broad and general attack family still deserves further exploration. 

Different types of privacy in ML are fragmental and overlapped, as concluded based on both previous empirical attacks and theoretical analyses. Without a solid understanding of privacy, TEE protection itself is not plausible; protection of one type of privacy could be invalid when the privacy goal changes. Therefore, a rigorous definition of privacy for specific use cases is the cornerstone for TEE protection.

\subsection{Partitioned ML execution for heterogeneous TEEs}
For now and in the foreseeable future, most TEEs on modern devices will be heterogeneous, and many will keep small considering the ubiquitousness of devices and the requirement of small TCBs.
It is desired to avoid complex monolithic ML frameworks; so we could evaluate and partition the solution with a focus on protecting the most critical ML computation or stages. 

The partition can be done on different scales. On a low level, most ML's computation forms with numeral calculations \eg~Multiply–accumulate operations, floating point operations, or matrix multiplication. Running a portion of these computations inside TEE is straightforward, but determining whether computation is not an easy task. First, this determination needs well-defined privacy as above mentioned, \eg~which type of privacy to protect. Second, it requires the knowledge of how these computations have different leakage of the target privacy. Currently, we have a superficial understanding of layer-level privacy differences, \ie~the last layer contains the most membership privacy. However, there is still a need i) to deepen the layer-level understanding of more types of privacy; ii) to investigate efficient layer-wise partitioning approaches; iii) to investigate any more fine-grained privacy understanding (\ie~matrix multiplication and feature maps), as the layer-level partition is still heavy for constrained TEEs.
Furthermore, one can also execute some kinds of assurance and verification inside TEEs instead of hiding sensitive parts directly. Such a method has been used for ML integrity checking~\cite{zhang2020enabling} but still deserves further investigation \ie~combining with masking techniques to preserve privacy.

\subsection{Dedicated TEE designs for general ML}
Efficient ML execution requires modern computer architectures with parallelization features, \eg~multi-threading on CPUs to load data and on GPUs (or TPU, etc.) to perform matrix multiplication in forward and backward passes. 
These features will need dedicated (co)designs of the hardware and the system of accelerated processing units, which most of the current TEEs do not have.
One solution is to equip TEEs with such parallelization features to increase the computational capability of ML. However, consequently, new features increase the TCB size. Synchronization bugs cause severe vulnerability of existing TEEs~\cite{weichbrodt2016asyncshock}, and one can also expect a considerable number of bugs that potentially exist in GPU-TEEs.
How to avoid such a dilemma really depends on how can we limit the trust boundary and reduce the TCB when applying parallel processing. 
One example is to consider one ML accelerated unit (GPU/TPU) to be physically isolated from the rest of the main board and the processing system in the same way as TPM, and to be accessed only through a secure bus from TEE-enabled CPUs. Although this constrains the trust boundary with assumptions that this dedicated GPU serves only for such private ML and no one can physically breach the GPU, there are more use cases applicable such as MLaaS with proper remote attestation and verification. Recent work on designing secure IO paths and attestation support for interactions between various confidential VMs (e.g., TDX, SNP-SEV, CCA) and confidential GPUs shows great progress toward trustworthy systems dedicated to ML workloads~\cite{nvidia2024blackwell, cheng2024intel}.

Interestingly, aside from acceleration via core processors, the arising technique called in-memory computing~\cite{sebastian2020memory, ielmini2018memory} offers another way for speedy data processing, also for ML~\cite{eleftheriou2019deep, wozniak2020deep, sebastian2020memory}. As an example, Mythic introduced a memory architecture for in-memory matrix multiplication. Compared to traditional computing architecture, it avoids a large amount of data transmitted across CPUs, SRAM, and DRAM, which could then speed up to 100$\times$ in some cases. This potentially leads to interesting interactions with the most recent TEE designs (SGXv2 and Arm CCA) that support large memory sizes.

\subsection{TEE-aware ML}
Due to the heterogeneity of devices' TEEs and ML workload, it is obvious that one solution cannot satisfy all needs. Therefore, designing the ML to be aware of TEEs and the capability of TEEs is critical.
To achieve this, ML frameworks will determine the maximum workload to be executed with TEEs which first requires the TEE's information such as memory size, processor speed, protected storage, and potentially other features like multi-threading and secure communication channels. Then, ML computations from the most sensitive to less are deployed into TEEs. Such a determination involves many variables that influence factors from system cost to ML performance; thus, one can potentially form a multi-objective optimization problem similar to Neural Architecture Search for resource-constrained environments~\cite{fedorov2019sparse, banbury2021micronets, sudharsan2021ml}. Furthermore, one may need techniques such as sparse approach~\cite{dai2022dispfl, li2021hermes} of knowledge distillation~\cite{diao2020heterofl,zhang2021parameterized} to handle the heterogeneity of TEEs and differences in model architectures, especially when updating models in FL.

\subsection{Protecting the full ML pipeline}
Protecting the full pipeline is intractable without the participation of multiple TEEs (\eg~\cite{volgushev2019conclave, mo2021ppfl}). Multiple TEEs are able to provide multiple trusted areas covering more devices and consequently more pieces of the pipeline. To achieve such ``multi-party computation'' based on TEEs, one needs to enable a verification mechanism for multiple participants so that TEEs from different chip manufacturers can be remotely attested by two or more challengers. For example, potentially one third-party proxy attestation service can help to verify TEEs from both Arm and Intel~\cite{veracruz} so that the TEE system could cover multiple locations of the ML pipeline. Also, we anticipate more usage of virtualization-based enclaves (such as Microsoft VBS enclaves~\cite{microsoft2024vbs}) on edge devices that do not have proper support for hardware enclaves. This approach introduces interesting opportunities for achieving more flexible and pluggable security models but also introduces new security challenges that are not fully explored yet.

In addition to the TEE workflow, determining the most sensitive parts of ML for protection is also not trivial. 
On the smallest scale, TEE provides a more trustworthy area in the ML pipeline, such as an additional trust base for random number producer or DP noise addition). Besides, investigations on the privacy or integrity of the pipeline's parts other than training/inference protection, such as data preparation, are vastly missing~\cite{singh2021quoc}. Such protection on one specific part of the pipeline will involve again threat (privacy/integrity) definition, sophisticated TEE-based protection design, and performance measurement similar to the training stage protection. After such works on more stages of the pipeline, full ML pipeline protection will be formed concretely and be used to a greater extent and larger scale.

\subsection{Risk of Side-channel Attacks}
Additionally, although it is out-of-scope in typical threat models of Confidential Computing, the risk of side-channel attacks would be increased due to the increasingly complex process when incorporating the ML process, which should be better analyzed and handled.
First of all, conducting side-channel attacks to disclose the structure of DNNs has been shown to be possible in various ways. Privado~\cite{grover2018privado} and Hua's work~\cite{hua2018reverse} demonstrated that by observing memory access patterns a CPU-TEE accesses, the adversary can infer the DNN's network structure. Ganred~\cite{liu2020ganred} and Cache Telepathy~\cite{yan2020cache} used only cache timing side-channel information and still were able to extract a targeting modern DNNs from a CPU-TEE. Similarly, electromagnetic (EM) emanations and power information, also can be leveraged to disclose DNN information~\cite{batina2019csi, yu2020deepem}.

Such vulnerability becomes severe with application-specific processors like CNN accelerators. For instance, Hermes~\cite{li2021hermes} observed the data stream on the PCIe bus between GPUs and CPUs, and Leaky DNN~\cite{wei2020leaky} monitored the resource usage of the GPU's kernel; both can steal DNN models' hyper-parameters, even identical weight information. The same side-channel issue also exists in microcontrollers that have been massively used in pervasive applications such as wearables and surveillance cameras shown by~\cite{batina2019csi}'s attack. These potential and new rising side-channel attacks should be delicately examined when implementing a Confidential Computing system for ML. Still, there are countermeasures that transform applications (\eg~ML frameworks) to be free of input/output-dependent access patterns by introducing obliviousness into the input/output. Though recent work shows progress in various mitigation techniques~\cite{oh2020trustore}, we are not even close to bridging this gap; hence we anticipate further developments tackling side-channel risks to emerge in the future.

\subsection{Comparative Analysis and Broader Impacts}
Liu et al.~\cite{liu2021machine} give a solid survey on privacy-related works with ML, including both ML for privacy and private ML. However, the survey considers only encryption-based protection, \ie~homomorphic encryption, multi-party computation, differential privacy, etc., and does not include any hardware-based protection mechanism such as TEEs. TEEs are becoming one of the prominent protection mechanisms, along with the act of GPU manufacturer NVIDIA, and are much more efficient than these encryption-based protections, which is our main target in this survey.
Note that in the previous work~\cite{liu2021machine}, attacks are grouped similarly to our categories but mostly following the name of the original work. The Model Extraction Attack refers to stealing the model, corresponding to model privacy; The Model Inversion Attack and Shadow Model Attack refer to inferring certain attributes or statistical properties of the training dataset, corresponding to AIA; The Model Memorization Attack refers to recovering the exact input data, corresponding to DRA. However, the original names are more or less inaccurate. For instance, the phrases ``Model Inversion Attack'' and ``Model Memorization Attack'' both implicitly express that the attack leverages the models' information to invert/reverse the information of others like training data, but are not distinguished on whether the information is exact data or features/attributes of the data. Our work aligns with more recent surveys on privacy attacks in ML~\cite{hu2022membership, salem2023sok, rigaki2023survey} and NIST's publication on the taxonomy and terminology of ML attacks~\cite{vassilev2024adversarial}, and we appeal for these more explicit taxonomies in future research.

Both previous surveys~\cite{duy2021confidential} and \cite{babar2023trusted} dedicatedly summarize Confidential Computing-assisted ML with detailed descriptions of both TEE and ML scenarios. 
However, they have not examined the complete ML pipeline; therefore, important components may be missing, such as i) the software stack that connects the TEE hardware stack to ML computation, and ii) the input data that is always outside of TEE, which, in contrast, has been comprehensively analyzed in our survey. 
Our work not only performs a more comprehensive summarization of previous work but also gives a ``bird's-eye view'' of the interdisciplinary to identify the necessary components to close the gap between Confidential Computing and ML. Consequently, our survey presents more insights into underlying challenges, limitations of existing systems, and potential solutions, which realize this work as a Systematization of Knowledge. 

From a broader perspective, ML with Confidential Computing could potentially provide solutions for many regulations such as GDPR~\cite{gdpr}, CCPA~\cite{ccpa}, and EU AI Act~\cite{euaiart}. As stated in the EU AI Act, existing EU law on the protection of privacy and confidentiality still applies to the use of information for AI-based technologies.
TEEs establish enclaves with restricted access for processing personal data, which consequently provides a level of privacy guarantees and confidentiality assurance, \ie~``the right to restrict processing'' in GDPR.
Interestingly, with the above-mentioned privacy measures (in Section~\ref{sec:privacyfoundation}), we gain a better understanding, both theoretically and empirically, of the level of information contained in the output of the TEE. This also brings forth GDPR's ``the right to explanation'' of the AI algorithm's outputs to a certain extent.
We believe with advanced techniques being developed in the future, Confidential Computing will be better integrated with ML and move forward in real-world applications while satisfying regulations.

%% file: 8-conclusion.tex
\section{Conclusion}
\label{sec:conclusion}

Protecting ML against privacy leakages and integrity breaches using Confidential Computing techniques is an exciting and challenging new era. 
Although many studies have been dedicated to running the training and inference processes inside TEEs, they still face the limitation of available trust resources. 
Since ML requires much more trusted resources; the current protection levels only provide the confidentiality and integrity of that specific stage, training/inference, in the complete ML pipeline. 
Current defenses against privacy leakage are usually measured through red-teaming and attack-specific metrics, which make the final results less reliable and generalized.
Furthermore, large attack surfaces exist at stages, especially at the upstream stage like data preparation, which can cause huge negative impacts on the ML pipeline. 
Rethinking how to tackle such a dilemma, \ie~the conflict between the large scale of the complete ML pipeline and the need for small TCB, is of utmost importance when conducting ML with Confidential Computing.

Confidential Computing achieves a hardware-based root-of-trust establishing a more trustworthy execution environment for ML activities, but we should rethink whether ``hiding'' the training/inference process inside such enclaves is the optimal solution. 
Future researchers/developers need to better understand the underlying privacy issues in the ML pipeline so that later protections focus on vital parts. 
The full ML pipeline needs to be evaluated to avoid fruitless labor on training/inference protection only. 
The current trends indicated that both TEE evolution and ML advances (\ie~partitioning ML execution for TEEs, and designing dedicated TEEs for ML) will lead to more \emph{TEE-aware} future architectures. 
These potential solutions can help achieve much stronger privacy and security guarantees without compromising ML computational performance and introducing TEE's system overhead.

\mysub{Suggestions} to achieve ML with Confidential Computing for the hardware/system/framework designer are:
\begin{itemize}[leftmargin=10pt]
    \item Compared to security or integrity, privacy definition is still ambiguous. For a targeted privacy guarantee goal, one needs to apply privacy definitions with strong theoretical foundations, such as differential privacy or $\mathcal{V}$-usable information.
    \item Protecting the upstream of the ML pipeline such as the data preparation is of utmost importance because its lack has intractable impacts on the downstream ML process. It can potentially be realized by integrating TEE-based verification into the data signature. Multiple TEEs/Conclaves can further benefit the full ML pipeline protection.
    \item Developing dedicated TEEs for general ML requires rethinking and redesigning the hardware acceleration features and corresponding system stack. New paradigms such as in-memory computing that involves a smaller group of hardware and software modules may ease the condition due to the reduction of the TCB size and the attack surface.
    \item Designing the ML framework to be TEE-aware and partitionable for heterogeneous TEEs requires i) meticulously researching the privacy/integrity weakness of different ML granularity (from layer/feature map to numeral calculations), and ii) managing the TEE system to protect the most sensitive ML components efficiently (\eg~with masking) with a high priority.
\end{itemize}